\documentclass[aps,prx,twocolumn]{revtex4}
\usepackage{graphicx}
\usepackage{amsmath}
\usepackage{braket}
\usepackage{hyperref}
\usepackage{subfigure}
\usepackage[usenames,dvipsnames]{color}

\usepackage{graphicx}
\usepackage{subfigure}

\newcommand{\ct}{\cite}

\newcommand{\bi}{\bibitem}
\newcommand{\be}{\begin{equation}}
\newcommand{\ee}{\end{equation}}
\newcommand{\ba}{\begin{eqnarray}}
\newcommand{\ea}{\end{eqnarray}}

\newcommand{\noi}{\noindent}

\begin{document}
\title{Interconnections between equilibrium topology and dynamical quantum phase transitions in a linearly ramped Haldane model}
\author{Utso Bhattacharya and Amit Dutta \\
Department of Physics, Indian Institute of Technology, 208016, Kanpur}

\begin{abstract}
{We study
the behavior of Fisher Zeros (FZs) and dynamical quantum phase transitions (DQPTs) for a linearly ramped Haldane model  
 occurring in the subsequent  temporal evolution  of the same and  probe the intimate connection with  the equilibrium topology of the model}. Here, we investigate the temporal evolution of the final state of the Haldane Hamiltonian (evolving with time-independent final Hamiltonian) reached 
following a linear ramping of the staggered (Semenoff) mass term from an initial to a final value, {first selecting a specific protocol}, so chosen that the system is ramped from one non-topological phase to the other through a topological phase.
{We establish the existence of three possible  behaviour of areas of FZs corresponding to a given sector: (i) no-DQPT, (ii) one-DQPT (intermediate) and (iii)two-DQPTs (re-entrant),  depending
on the inverse quenching rate $\tau$.}
Our study also reveals that the appearance of the areas of FZs is an artefact of the non-zero (quasi-momentum dependent) Haldane mass ($M_H$), whose absence leads to an emergent one-dimensional behaviour indicated by the shrinking of the areas FZs to lines and the non-analyticity in the dynamical "free energy" itself. Moreover, the characteristic rates of crossover between the three behaviour of FZs are determined by the time-reversal invariant quasi-momentum points  of the Brillouin zone where $M_H$ vanishes. Thus, we observe that through the presence or absence of $M_H$, there exists an intimate relation to the topological properties of the equilibrium model even when the ramp drives the system far away from equilibrium.

\end{abstract}
\maketitle

\section{Introduction}
The investigation of non-equilibrium many body quantum systems has always been a challenging task, both experimentally and theoretically.   The recent experimental advances in  realisation of closed condensed matter systems via cold atoms \ct{bloch08,lewenstein12} in optical lattices have paved the way  for addressing unresolved fundamental problems in out of equilibrium strongly correlated systems. Henceforth, experimental  studies  have been performed to probe intriguing dynamical phenomena like the real time evolution of closed quantum systems in cold atomic gases, \ct{greiner02},   prethermalization \ct{kinoshita06,gring12,trotzky12},  light-cone like propagation of quantum
correlations \ct{cheneau12},  light-induced non-equilibrium  superconductivity and topological systems \ct{fausti11,rechtsman13} and many-body localization in disordered interacting systems \ct{schreiber15}. In parallel, there have been a plethora of theoretical works on e.g., the growth of entanglement entropy following a quench \ct{calabrese06}, thermalization \ct{rigol08}, light-induced topological 
matters \ct{oka09,kitagawa10,lindner11}, dynamics of topologically ordered systems \ct{bermudez09,patel13,thakurathi13}, periodically driven closed
quantum systems \ct{mukherjee09,das10,Russomanno_PRL12,bukov16,sen16} and many body localization \ct{pal10,nandkishore15}, to name a few.
{(For review, we refer to \ct{dziarmaga10,polkovnikov11,dutta15,eisert15,alessio16,jstat}.)}

 One of the emerging features associated with a non-equilibrium quantum many body system  is the possibility of dynamical quantum phase transitions in quenched closed quantum systems, introduced by Heyl $et~ al$ \ct{heyl13}: here, non-analytic behavior occurs at critical times in the subsequent real time evolution (following
 the quench)
 generated by the time independent final Hamiltonian.
  To probe DQPTs in a closed quantum system, one prepares a desired final state $\ket{\psi_f}$ of the system by quenching (changing slowly or rapidly) the parameters of the Hamiltonian  usually from one phase to another across a quantum critical point (QCP) \ct{sachdev96}. Then, this state is allowed to freely evolve in time with the time-independent final Hamiltonian $\left(H_f\right)$ 
 up to a time $t$ (measured from the instant when $\ket{\psi_f}$ is prepared), before taking its overlap with the final state at the start of the free evolution. This so called  Loschmidt overlap amplitude (LOA) is denoted by, $G(t)=\braket{\psi_f|e^{-iH_ft}|\psi_f}$, where the final state is not an eigenstate $H_f$. (It should be noted  that throughout this paper  the Planck constant $\hbar$ is set equal to unity.)
Generalization  to the complex  time ($z$) plane yields $G(z) = \braket{\psi_f|e^{-H_fz}|\psi_f}$ where $z = {\rm Re}[z] +it$.  Drawing a  formal analogy between the  inverse temperature ($\beta$) and complex time ($z$) \ct{fisher65,lee52,saarloos84}, we notice that there exists a close similarity between the canonical partition function of an equilibrium classical system $Z(\beta )={\rm tr}\left(\exp({-\beta H})\right)$ and $G(z)$ which can now be referred to as a dynamical partition function. By extending this analogy with equilibrium classical phase transitions further, one can now define a quantity in the thermodynamic limit for a $d$-dimensional system with linear dimension $L$, called the dynamical free energy density 
$$f(z)= -\lim_{L\to\infty}\frac{1}{L^d}\ln{G(z)}.$$ 
One can then similarly look for zeros of $G(z)$ (equivalently, non-analyticities in $f(z)$)  to find the so-called ``Fisher zeros" (FZs) residing in the complex $z$ plane. 
 These zeros of the dynamical partition function $\left(G(z)\right)$ cover lines or areas (more specifically closely spaced points for a finite size systems which form
 continuous lines or dense areas in the thermodynamic limit) in the complex time ($z$) plane (depending upon the {\it effective} dimensionality of the system under consideration). 
{ Consequently, when  the lines (or areas) of FZs  cross  the imaginary (real time) axis  at dynamical critical (real)  times $t_c = {\rm Im}[z_c]$, one observes DQPTs  manifested in 
 non-analyticities in Re[$f(t)$] (or its time derivative  Re[$f'(t)$]).  Evidently,  the LOA, i.e., $G(z)$ can only decay to zero, when the prepared many-body final state $\ket{\psi_f}$  becomes orthogonal to its time evolved counterpart after free evolution with $H_f$.}

 The occurrence of DQPTs   at specific instants of time  following a rapid quench of a   transverse Ising chain \ct{suzuki13} across its  QCP was  established in Ref. [\onlinecite{heyl13}] and the lines of FZs were indeed found to cross the real time axis at those instants. This observation  has been independently confirmed through several  works on quenched  one-dimensional (1D)  integrable
and non-integrable systems \ct{karrasch13,kriel14,andraschko14,canovi14,heyl14,heyl15,palami15,divakaran16,huang16,puskarov16,zhang16,heyl16,zunkovic16,zvyagin17,sei17,fogarty17}.  Subsequent studies, however, have established that sudden quenching within the same phase of a system (both integrable and non-integrable) without ever encountering an equilibrium QCP may still give rise to DQPTs in some situations \ct{vajna14,sharma15}.  In an attempt to 
characterise a DQPT  through an order parameter, the crucial role played by topology has been exploited to define a dynamical topological order parameter (DTOP); the
DTOP changes its integer value whenever the system dynamically crosses a critical time $t_c$  signalling the occurrence of a DQPT \ct{budich15}. 

{DQPTs have also been observed for one dimensional systems  when the state $\ket{\psi_f}$ is prepared through slow quenching of a parameter of the
initial Hamiltonian across QCP(s) \ct{pollmann10,sharma16}. Unlike the sudden quenching case, for slow quenching DQPTs survive in the subsequent evolution  following quenches across two QCPs; moreover, the lines of FZs form a lobe like structure
in the complex $z$ plane which is also reflected in the temporal evolution of the DTOPs \ct{sharma16}. We note in passing that the study of a slow ramping across a QCP
have gained importance because of the possibility of universal Kibble-Zurek scaling \ct{kibble76,zurek85}  of the defect density and the residual energy in
the final state reached after the quench \ct{zurek05,polkovnikov05} (for review see, [\onlinecite{polkovnikov11,dutta15,dziarmaga10}]).} 
Generalising  to two dimensions, the possibility of the occurrence of DQPTs following  a sudden quench from the non-topological phase to the topological phase of the Haldane model \ct{vajna15} and also from the gapped  to the gapless phase of a Kitaev honeycomb model have been explored \ct{schmitt15}. These studies  have shown that in stark contrast to that of the 1D case, the FZs  cover areas instead of lines.

We further note that Re$[f(t)]$ i.e., $(-1/L^d) {\rm Re}[\ln \braket{\psi_f|e^{-iH_ft}|\psi_f}]$, is closely related to the  Loschmidt echo which has
been studied extensively close to a QCP both in equilibrium \cite{quan06,rossini07,cucchietti07,sharma12} and non-equilibrium situations \ct{venuti10,mukherjee12,nag12,dora13} and
shares a close connection with several other aspects of non-equilibrium dynamics of quenched closed quantum system \ct{gambassi11,russomanno15,zanardi07_echo,dorner12,sharma15_PRE}. 
Re$[f(t)]$  can also be
interpreted as
 the rate function of the return probability, so called because it can be connected to the singularities in the work distribution function corresponding to zero work following a double quenching process \ct{heyl13}.  

Very recently, Flaschner {\it et al.}, \ct{flaschner16} has reported the first experimental observation of a DQPT using time-resolved state tomography to determine the dynamical evolution of a fermionic many-body state after a quench between two lattice Hamiltonians  {as shall be discussed later in this work.}
  {Furthermore, DQPTs have also been investigated in a string of ions simulating interacting transverse-field Ising models  in the  non-equilibrium dynamics induced by a quantum quench \ct{jurcevic16}: these transitions have been measured  through a quantity that becomes non-analytic in time in the thermodynamic limit.}

Consequently, these experiments motivate us, in this paper, to address the possibility of the occurrence of DQPTs following a   {slow  quench of  the paradigmatic Haldane model on a hexagonal lattice \ct{haldane88} and investigate the connection between the equilibrium topology of the same model  and DQPTs following a slow ramping. We note that the Haldane model
has been  experimentally realized \ct{jotzu14}  and for long it has been providing the theoretical base for exploring topological insulators.  {(In a later experiment,
the Chern number of   Hofstadter bands with ultracold bosonic atoms was measured \ct{aidelsburger15}.)} We would like to emphasise at the outset that the focus of this paper is strictly restricted
to exploring the role of topology on  DQPTs occurring in a quenched Haldane model only and hence, we shall abstain from drawing any parallel to  other two dimensional topological or  non-topological 
models.   
 To the best of our knowledge, this is the first attempt to explore the DQPTs following {\it slow quenches} in a higher dimensional (equilibrium)  system which
 may exhibit effective one-dimensionality in limiting situations. In this work,  the (staggered) Semenoff mass  term \ct{semenoff08}  in the topological Haldane model is slowly ramped via a linear quenching protocol with an inverse  quenching rate $\tau$ so that the system {is quenched from one non-topological phase to the other across the topological
 phase in the process as shown in Fig.~\ref{fig_haldane_phase}:}
 To observe the DQPTs, one then tracks the LOA of the final
 state evolving with the time-independent final Hamiltonian. Our study unearths the vital roles played  by the parameter $\tau$  and more fundamentally, by the
 Haldane mass in dictating the behavior of FZs  and consequently in resultant DQPTs. We note that the slow ramping of the Haldane
 model has also been studied in the context of inducing topological transitions \ct{privitera16}. }

  {Let us summarize our main   {observations} at the outset {initially based on the numerical studies of a particular quenching protocol between the same initial and final points, depicted in Fig.~\ref{fig_haldane_phase}: depending on the inverse quenching rate $\tau$, we observe three distinct behavior of areas of FZs corresponding to
 a single sector. For very slow ramping, if the rate $\tau$ exceeds a critical value, the area of FZs  crosses  the imaginary $z$-axis (real time axis) twice.
 As a consequence, there exist
 four boundary points of the area
those cut the real time axis resulting in four instants of real time where the first derivative of the dynamical free energy shows cusp singularities. For a very rapid quench, on the other hand,
the areas do not cross the real axis and there is no DQPT.
 {Notably, we also find the} existence of an intermediate
 range of $\tau$, separating the no-DQPT region  from the re-entrant region,  for which the area of FZ  cuts the real axis once; as a result,  there are  only two
 instants of non-analyticities. }
 Most importantly, the (quasi-momentum dependent) Haldane
 mass ($M_H(\mathbf{k})$), which makes the Haldane model topologically non-trivial in  the phases with non-zero Chern numbers, plays a more fundamental role:  \ {$M_H(\mathbf{k})$  is  anisotropic
 in the sense that it is an explicit function of $k_x$ and $k_y$ (  {i.e., not a function of $|k|$ only})  and consequently it leads to  the areas of FZs (2D behavior)}. Furthermore, the critical values
 of $\tau$ (which dictate the cross-over from the no-DQPT region to the intermediate and from the intermediate to the re-entrant region) are determined by the
 time reversal invariant momentum (TRIM) points  of the Brillioun Zone (BZ) (as depicted in Fig.~\ref{fig_BZ}) for which $M_H(\mathbf{k})=0$.  
 In the situation, when the Haldane mass is altogether absent 
 in the equilibrium model
 the FZs {exhibit an emergent 1D behavior} and form lines ({\it not} areas) which cut the real time
 axis at specific instants of time leading to non-analyticities in Re$[f(t)]$ itself: 
 In short, the Haldane mass induces a {\it dimensional crossover } 
 in the context of DQPTs. 
   { In the concluding section, we shall summarise which of the above observations are generic and would hold true  for any quenching
 protocol. }
 
 
 {Let us now elaborate on the organization of the paper. In Sec. \ref{sec_haldane}, we introduce the Haldane model along with  the quenching protocol  emphasizing the role of  the Haldane mass. In Sec. \ref{sec_results} we present and analyze our main results illustrating the behavior of excitation probabilities and
 the areas of FZs for different values of $\tau$.  In Sec. \ref{sec_zero_MH}, we critically analyze
 the role of $M_H(\mathbf{k})$ and illustrate how a dense area of FZs emerges from disjoint lines  {while in Sec. \ref{sec_expt}, we discuss the essential experimental
 connections.}
 Other than concluding comments presented in Sec.~\ref{sec_conclusion},
   we further include one Appendix, 
 where a brief note on the Haldane model is provided.
 
 \section{Haldane Model and the quenching scheme}
 
 \label{sec_haldane}

\label{sec_haldane}
Let us consider the 2D topological Haldane model on a hexagonal lattice comprised of two triangular sublattices $A$ and $B$ {(refer to Appendix \ref{app_model} for detail)}. The Haldane Hamiltonian is based on a graphene-like Hamiltonian but with a sublattice symmetry (SLS) breaking  Semenoff mass (SM)  term ($M$) and  a staggered magnetic field which manifests itself in the complex next nearest neighbour hopping. The presence of a periodic boundary condition enables us to write down the Hamiltonian in the quasi-momentum $\left(k_x,k_y\right)$ basis as, 

\begin{widetext}
\begin{equation}
\begin{aligned}
\mathcal{H}(\mathbf{k}) & = \begin{pmatrix}
c^{\dagger}_A(\mathbf{k}) & c^{\dagger}_B(\mathbf{k}) 
\end{pmatrix}
\begin{pmatrix}
h_3(\mathbf{k}) & h_1(\mathbf{k})+ih_2(\mathbf{k}) \\
h_1(\mathbf{k})-ih_2(\mathbf{k}) & -h_3(\mathbf{k})
\end{pmatrix}
\begin{pmatrix}
c_A(\mathbf{k}) \\
c_B(\mathbf{k}) 
\end{pmatrix},
\end{aligned}
\label{eq_haldane_model}
\end{equation}

\noi where $h_3(\textbf{k})= M + M_H(\mathbf{k})$ with $M_H(\mathbf{k})$ being the quasi-momentum dependent Haldane mass given by
\begin{equation}
M_H(\mathbf{k}) = 2 t_2 \sin (\phi) \biggl[\sin  \left(\mathbf{k} \cdot \mathbf{a}_2 \right) - \sin \left( \mathbf{k} \cdot \mathbf{a}_1 \right)+ \sin \left( \mathbf{k} \cdot \left(\mathbf{a}_1 - \mathbf{a}_2 \right) \right)\biggr],
\label{eq_haldane_mass}
\end{equation}
\end{widetext}
\noi that vanishes for $t_2=0$ or $\phi=0$, while 
\begin{align}
h_1 &= t_1 \biggl[ 1 + \cos \left( \mathbf{k} \cdot \mathbf{a}_1\right) + \cos \left(\mathbf{k} \cdot  \mathbf{a}_2 \right) \biggr]\nonumber\\
h_2  &= t_1   \biggl[\sin  \left( \mathbf{k} \cdot \mathbf{a}_1 \right) +\sin \left( \mathbf{k} \cdot\mathbf{a}_2 \right) \biggr],\\
\end{align}
with $\mathbf{a}_1 =\frac{a}{2} \left(3,\sqrt{3}\right)$ and $\mathbf{a}_2 =\frac{a}{2}  \left(-3,\sqrt{3}\right)) $.

\begin{figure*}[]
\centering
\subfigure[]{%
\includegraphics[width=.45\textwidth,height=6cm]{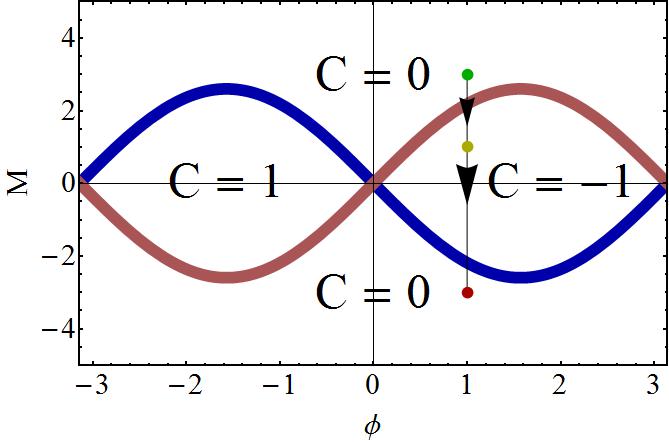}
\label{fig_haldane_phase}}
\hfill
\quad
\subfigure[]{%
\includegraphics[width=.45\textwidth,height =6cm]{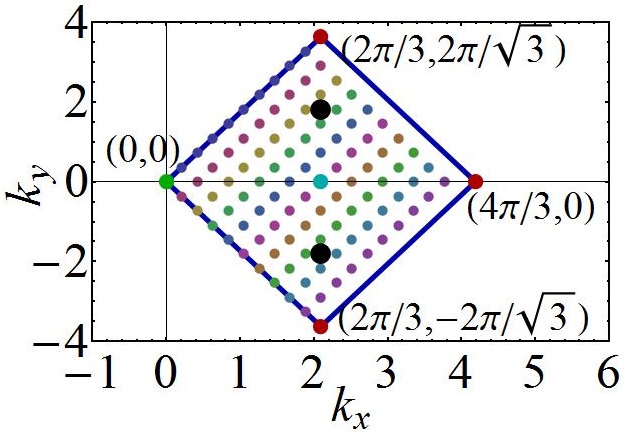}
\label{fig_BZ}}
\caption{(a) (Color online) The phase diagram of the Haldane model in $M-\phi$ plane; topological phases and the non-topological phase correspond to Chern numbers $C=\pm 1$,
 $0$, respectively. We employ a quenching scheme $M(t)=-t/\tau$ with $M_i=3$ (green dot) and $M_f=-3$ (red dot) with $\phi=1$, so that the system is quenched across both the critical lines (i.e., gapless DPs) in the
process of quenching. 
(One may also  consider a slow ramping from $M_i=3$ (green dot) to $M_f=1$ (yellow dot).)
(b) The rhomboidal Brillouin zone (BZ) that we
have chosen in this paper is illustrated for a $10 \times 10$ system. The Dirac points are shown in black (biggest dots) while other small dots represent the points
of the BZ. The four corner points and the highest symmetry point at the centre (with co-ordinates $(2\pi/3,0)$) marked by dots of medium size  are time reversal invariant momentum (TRIM) points; while the left corner point $(0,0)$ (in green) and the central point (in cyan) are included in the BZ, other TRIM points (in red) are not.   {At all TRIM points the
Haldane mass is zero.} }
\end{figure*}

We shall henceforth set the lattice constant $a=1$ and use a rhomboidal BZ as shown in Fig. \ref{fig_BZ}.
The time reversal symmetry (TRS) in this model is broken by the phase factor $\phi$ in the Haldane mass term, originating from the staggered magnetic field and is positive for anticlockwise nearest neighbour hopping. The breaking of TRS indicates that the two Dirac points ($\mathbf{K}$ and $\mathbf{K'}$) in bare graphene spectrum are no longer time-reversed partners of each other as each of them sees a different Haldane mass $M_H(\mathbf{k})$ depending on their quasi-momentum values, i.e., $\mathbf{K}$ or $\mathbf{K'}$. We will later see that  this  (Haldane) mass term depends on $k_x$ and $k_y$ explicitly (  {i.e., does not depend on $|k|$ only)}, as shown in Fig.~\ref{fig_haldane_mass}, and  is essential for  FZs to cover areas in
the complex $z$ plane.  {For a note  on the model \eqref{eq_haldane_model}, we refer to the Appendix \ref{app_model}.}\\

We now perform a slow quench on the Haldane model, initially in the ground state $\ket{1_i}$ of the initial Hamiltonian ${\mathcal H}_i(\mathbf{k})$, by linearly ramping the SLS breaking quasi-momentum independent SM  term $(M)$ using the protocol $M(t) = -{t}/{\tau}$  from an initial value $M_i=3$ to final value $M_f= -3$ with 
$\phi$ fixed to one as illustrated in Fig. \ref{fig_haldane_phase}. However, the results presented here would be in general true  except for the special situations when $\phi=0$ and $|M_i|, |M_f| \to \infty$. {In the former case, there is no non-trivial topology of the equilibrium model while the latter
situation refers to the infinite time Landau-Zener problem \ct{landau,sei} where the role of  topology is completely wiped out. In both the cases, we  thus arrive at  the problem of
analyzing  DQPTs in a linearly ramped gapped (Semenoff) graphene-like system as  elaborated in Sec. \ref{sec_zero_MH}. Although, the finite time
LZ problem can be studied within an analytical framework \ct{vitanov}, the results are not useful in the present context and consequently, we shall base our analyses and
inferences on extensive numerical calculations of the finite time LZ problem. \ {All these numerical results are analysed and explained using arguments
based on symmetry and topology.   {For the subsequent discussion, we present results for a particular quenching scheme and draw conclusions from
them. In the concluding sections, we shall summarise which of  these arguments would hold true for the generic situations.}}

 In the process of ramping, the system passes through two gapless critical lines (see Fig. \ref{fig_haldane_phase}) where the
 characteristic time scale (i.e., the relaxation time) diverges and hence,
 the condition for an adiabatic dynamics breaks down in their vicinity.  One arrives at a final state (for the $\mathbf{k}$-th mode) at the end of
 the quenching: 

\begin{equation}
\ket{\psi_f(\mathbf{k})}=u_f(\mathbf{k})\ket{1_f} + v_f(\mathbf{k})\ket{2_f},
\label{eq_final_psi}
\end{equation}

\noi where $|u_f(\mathbf{k})|^2+|v_f(\mathbf{k})|^2=1$, $\ket{1_f}$ and $\ket{2_f}$ are the ground and excited states  of the final Hamiltonian ${\mathcal H}(\mathbf{k}, M_f)$ with energy eigenvalues $e^1_f(\mathbf{k})$ and $e^2_f(\mathbf{k})$, respectively; {evidently, $|v_{\mathbf{k}}|^2$ stands for the probability of excitation  {following the ramp} and is denoted by $p_{\mathbf{k}}$ in the subsequent discussion.}

Thus, the role of the slow quenching process is to prepare the system in the desired final state, $\ket{\psi_f(\mathbf{k})}$ {which then evolves} in time with the final Hamiltonian $\left(\mathcal{H}_f(\mathbf{k})\right)$ yielding $e^{-    i \mathcal{H}_f(\mathbf{k})t'}\ket{\psi_f(\mathbf{k})}$, where $t'=t-t_f$ is measured from the instant when the final state is attained.
{One then immediately finds the LOA}, $G(t')=\prod_\mathbf{k}\braket{\psi_f(\mathbf{k})|e^{-i{\cal H}_f(\mathbf{k})t'}|\psi_f(\mathbf{k})}$, which contains contribution from all the quasi-momenta ($\mathbf{k}$) modes. 
Thus, to clarify, there are two kinds of evolution,  one that takes $\ket{1_i}$ at initial time $t_i = 0$ to final quenching time $t = t_f$  using the protocol
$M(t) = -t/\tau$, to reach the state $\ket{\psi_f}$, which then evolves with time-independent ${\cal H}_f(\mathbf{k})$.

\begin{figure}
\includegraphics[width=0.49\textwidth,height=7cm]{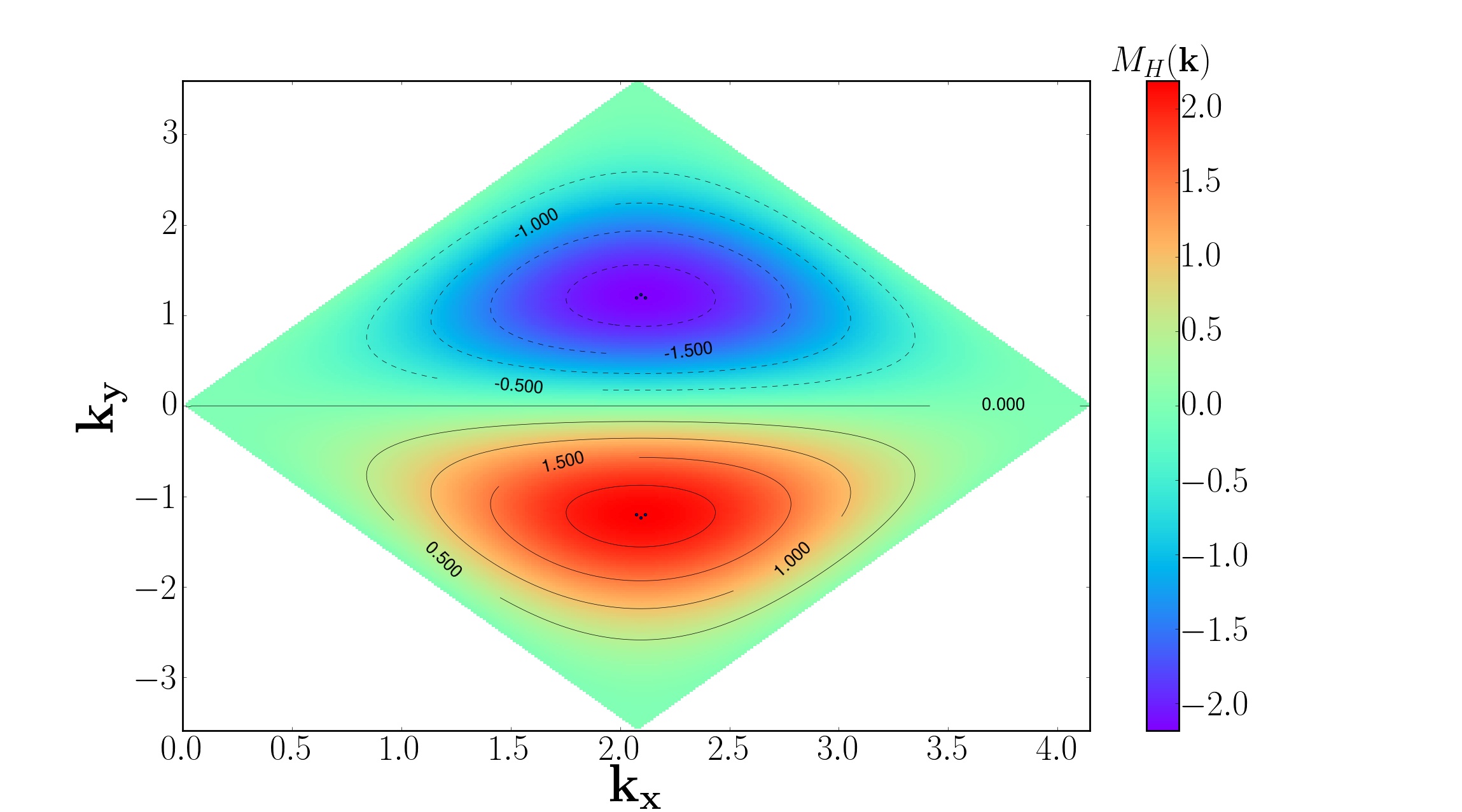}
\caption{The variation of the Haldane mass $M_H(\mathbf{k})$ in the Brillouin zone (BZ) as a function of $k_x$ and $k_y$. It  is zero along the $k_x$ axis 
when $k_y=0$ while it is  positive (negative) in the lower (upper)
half-plane. As a consequence of the
time reversal invariance of this mode, $M_H$ vanishes at all the TRIM points.  Following any contour of fixed $M_H(\mathbf{k})$ in the BZ, one can immediately  conclude that the Haldane mass must explicitly depend on $k_x$ and $k_y$   {and not $|k|$ only}.}
\label{fig_haldane_mass}
\end{figure}

{Moving on to the complex $z$-plane, we now solve for the zeros of the dynamical partition function (FZs)  $\left(G(z)=0\right)$ to locate the dynamical critical points. It is
then straightforward to show that  the FZs are given by,
\begin{equation}
z_n(\mathbf{k})=\frac{1}{e^2_f(\mathbf{k})-e^1_f(\mathbf{k})}\left[\log{\left(\frac{p_\mathbf{k}}{1-p_\mathbf{k}}\right)}\right]+i\pi(2n+1)]
\label{eq_fisher_zero}
\end{equation} 
where $n=0, \pm 1, \pm 2,  \cdots$ are integers;  a particular value of $n$ corresponds to one set of FZs and we shall essentially focus on the
$n=0$ sector.  {It should however be noted that  although the equation for FZs in Eq.~\eqref{eq_fisher_zero} have similar form as in the sudden quenching quenching case \ct{heyl13,vajna15}, the 
excitations probabilities  $p_\mathbf{k}$ in the present case are determined by the ramping protocol. We reiterate that these excitation probabilities are numerically estimated through a finite LZ problem for each mode $\mathbf{k}$.}  The  particle-hole symmetric nature of the Haldane Hamiltonian \eqref{eq_haldane_model} demands that   $e^2_f(\mathbf{k})=-e^1_f(\mathbf{k})$, yielding
\begin{equation}
z_n(\mathbf{k})=\frac{1}{2e^2_f(\mathbf{k})}\left[\log{\left(\frac{p_\mathbf{k}}{1-p_\mathbf{k}}\right)}\right]+i\pi(2n+1)].
\label{eq_fisher_zero1}
\end{equation}
}

\begin{figure}[]
\centering
\subfigure[]{%
\includegraphics[width=.45\textwidth,height=6cm]{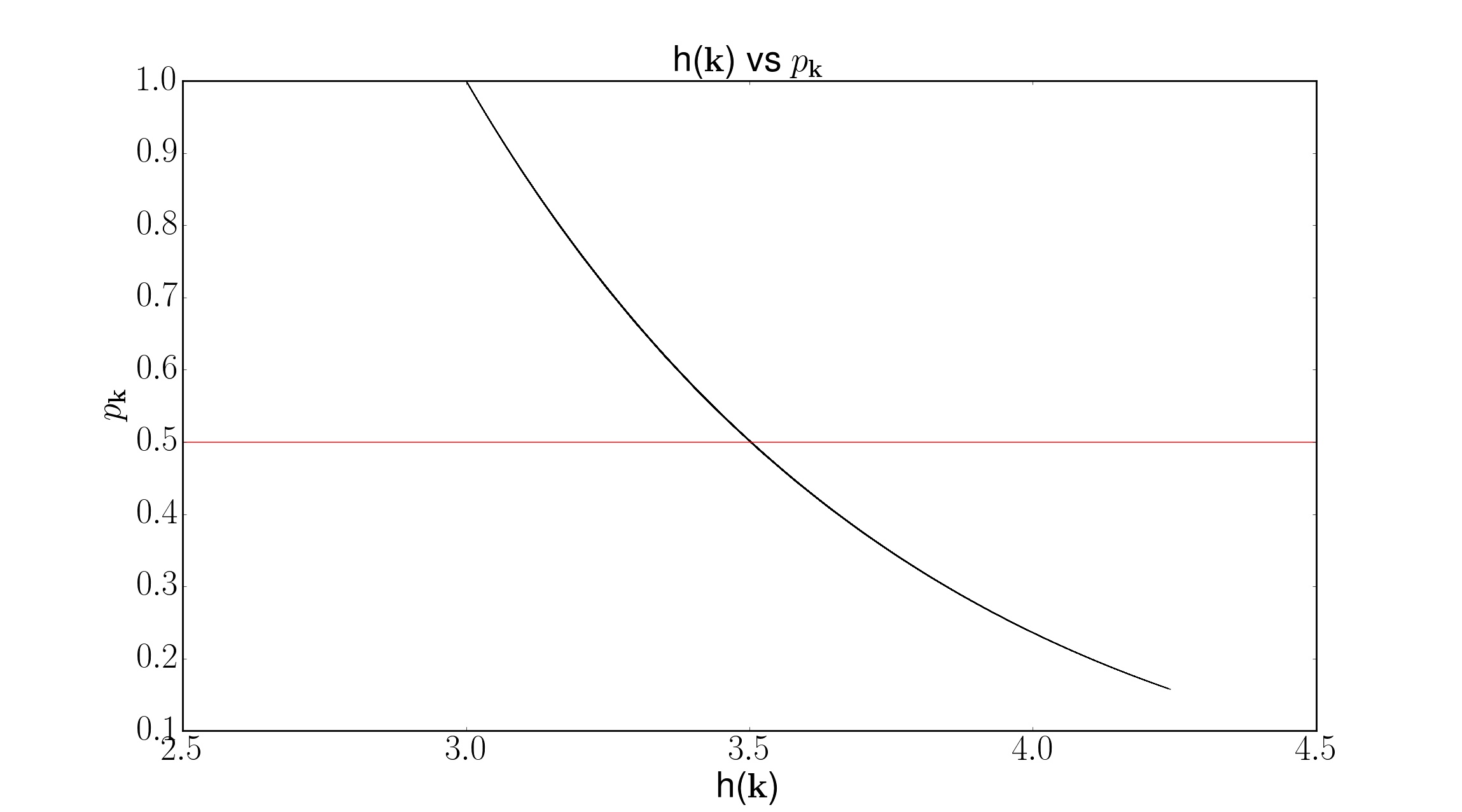}
\label{fig_resub1}}
\hfill
\quad
\subfigure[]{%
\includegraphics[width=.45\textwidth,height =6cm]{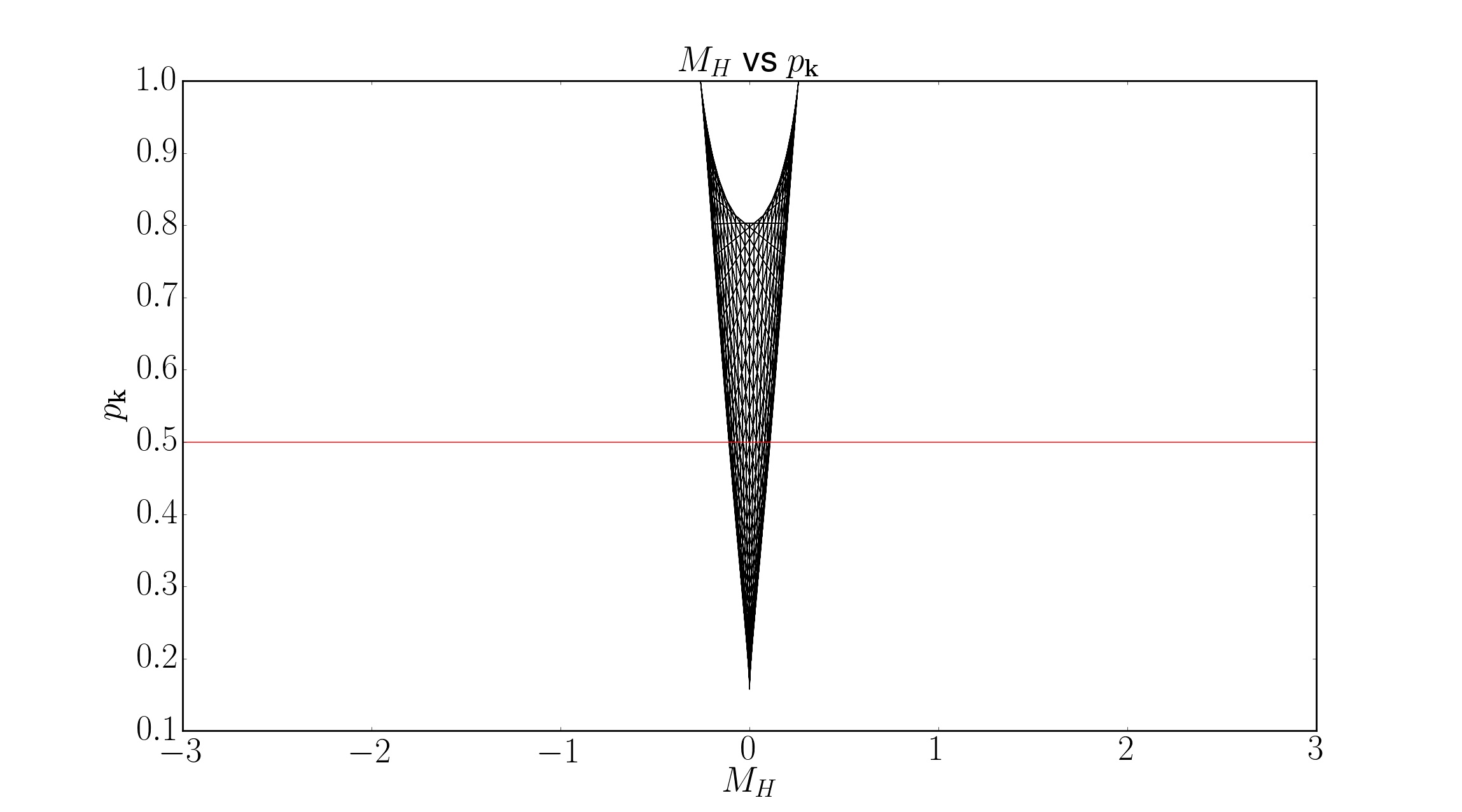}
\label{fig_resub2}}
\caption{(Color online) (a) The variation of $p_\mathbf{k}$ with $h(\mathbf{k})=\sqrt{h_1^2(\mathbf{k}) + h_2^2(\mathbf{k})+ M_f^2}$,  has been plotted for a quench from $M_i=3$ to $M_f=-3$ with a $\tau=0.1$ for a small $\phi=0.1$. The $p_{\mathbf{k^*}}=1/2$ (in red) line intersects with the curve at only one point indicating that all $\mathbf{k^*}$'s lead to the same value of $h(\mathbf{k^*})$. 
(b) The multiple intersections of the $p_{\mathbf{k^*}}=1/2$ line in the plot of $p_\mathbf{k}$ vs $M_H$ shows that  different $\mathbf{k^*}$s yield different contributions to $e^2_f(\mathbf{k^*})$ rendering it anisotropic, thereby leading to a continuum of critical $t_c$'s.}
\label{fig_resub}
\end{figure}

These FZs when plotted {in the complex $z$-plane} for each value of $k_x$ and $k_y$ (for a fixed $n$) may come together to form a line  or cover a dense area depending upon the effective dimension of the system. Substituting $p_\mathbf{k} = {1}/{2}$ for the mode $\mathbf{k} = \mathbf{k^*}$, in Eq.~ \eqref{eq_fisher_zero}  renders the real part of $z_n$ zero while the imaginary part is given by ${i\pi(2n+1)}/{2e^2_f(\mathbf{k^*})}=it_c$; at these instants of real time,
\be
t_c^{(n)}=\frac{\pi(n+\frac{1}{2})}{e^2_f(\mathbf{k^*})}
\label{eq_time}
\ee
one observes DQPTs, namely the non-analyticities (cusp singularities)  in ${\rm Re}{(f(t)}$ or ${\rm Re}(f'(t))$ with
 the critical times ($t_c^{(n)}$) being  inversely proportional to the energy (of the excited state) $e^2_f(\mathbf{k^*})$ of $H_f$. We also note that given the two-level nature of the problem
the condition $p_\mathbf{k^*} = {1}/{2}$ implies an effective infinite temperature state of the final Hamiltonian for the mode $\mathbf{k^*}$. 
For brevity, we shall henceforth drop the subscript $f$ in $e^2_f(\mathbf{k}$).

{Let us now illustrate how the presence of a non-zero $M_H(\mathbf{k})$ is essential in resulting in  dense areas of FZs in the complex $z$ plane. Given the two-dimensional nature
of the Haldane Hamiltonian, one naturally expects that the values 
of $k_x, k_y$  
with  the same  excitation probability  $p_{\mathbf{k}}$  must  lie within a continuum band or range. \ {It is now important to analyze whether these particular values of $\mathbf{k}$ for the entire range  correspond to the same value of $e^2(\mathbf{k})$; otherwise, Eq.~\eqref{eq_fisher_zero1} (for
a given $n$) ensures that the FZs corresponding to all these modes must  lie at different points in the complex $z$-plane.  }

\ {Referring to the Hamiltonian \eqref{eq_haldane_model},
we find that $e^2(\mathbf{k}) = \sqrt{(h_1^2(\mathbf{k}) + h_2^2(\mathbf{k})+[M_f + M_H(\mathbf{k})]^2}$. We note that the contribution to energy $e^2(\mathbf{k})$ from $(h_1^2(\mathbf{k^*}) + 
h_2^2(\mathbf{k^*})+M_f^2)$ for the values of $\mathbf{k^*}$'s for which $p_{\mathbf{k^*}}=1/2$ are all the same as is apparent from the Fig. \ref{fig_resub1}; this figure   shows that the ``$p_{\mathbf{k^*}}=1/2$"  line intersects only at one point with the $p_{\mathbf{k}}$ vs $h({\mathbf{k}}) =\sqrt{(h_1^2(\mathbf{k}) + h_2^2(\mathbf{k})+ M_f^2})$  curve. Consequently,  it turns out to be  absolutely necessary to explore the
variation of $M_H$  across the BZ   to conclude  about  the functional dependence of $e^2(\mathbf{k})$ on $k_x$ and $k_y$. From  Fig.~\ref{fig_haldane_mass},  we immediately conclude that  $M_H(\mathbf{k})$ is anisotropic in $k_x$ and $k_y$ and has  an explicit  functional dependence on them.  Now focussing on Fig. \ref{fig_resub2}, we observe that even when the Haldane mass is quite small with $\phi = 0.1$,  we see that the $p_{\mathbf{k^*}}=1/2$ line passes through a set of points corresponding to different values of $M_H(\mathbf{k^*})$. Therefore, such a dependence makes $e^2(\mathbf{k})$ depend explicitly on $k_x$ and $k_y$ and hence FZs with different $\mathbf{k}$ (corresponding to  the same $p_{\mathbf{k}}$) would lie at different points in the complex $z$-plane, resulting in an
area of FZs for a particular $n$. Further, from  Eq.~\eqref{eq_time}, we then
conclude that FZs corresponding to   different $\mathbf{k^*}$ must touch the real {time axis (imaginary $z$-axis)} at different instants.   Although the result presented in Fig. \ref{fig_resub} is plotted for a particular quenching scheme ensuring that ${\mathbf{k^*}}$ exists,  this needs to be emphasised that
the fact that the anisotropic continuous variation of the  Haldane mass is at the root of generating areas in the complex $z$-plane holds true  for any quenching protocol (sudden or slow) and any set of parameter values and  even  when  $\mathbf{k^*}$ may not exist}.
 On the contrary, when $M_H(\mathbf{k})=0$ in the equilibrium model, this
explicit dependence on $k_x$ and $k_y$ disappears, \ {as is evident from Fig.~\ref{fig_resub1},} and one gets an effective one-dimensional behavior where one observes lines of FZs, ({\it not} areas).   This issue
will be further elaborated in Sec. \ref{sec_zero_MH}.

\begin{figure*}[]
\centering
\subfigure[]{%
\includegraphics[width=0.49\textwidth,height=6.5cm]{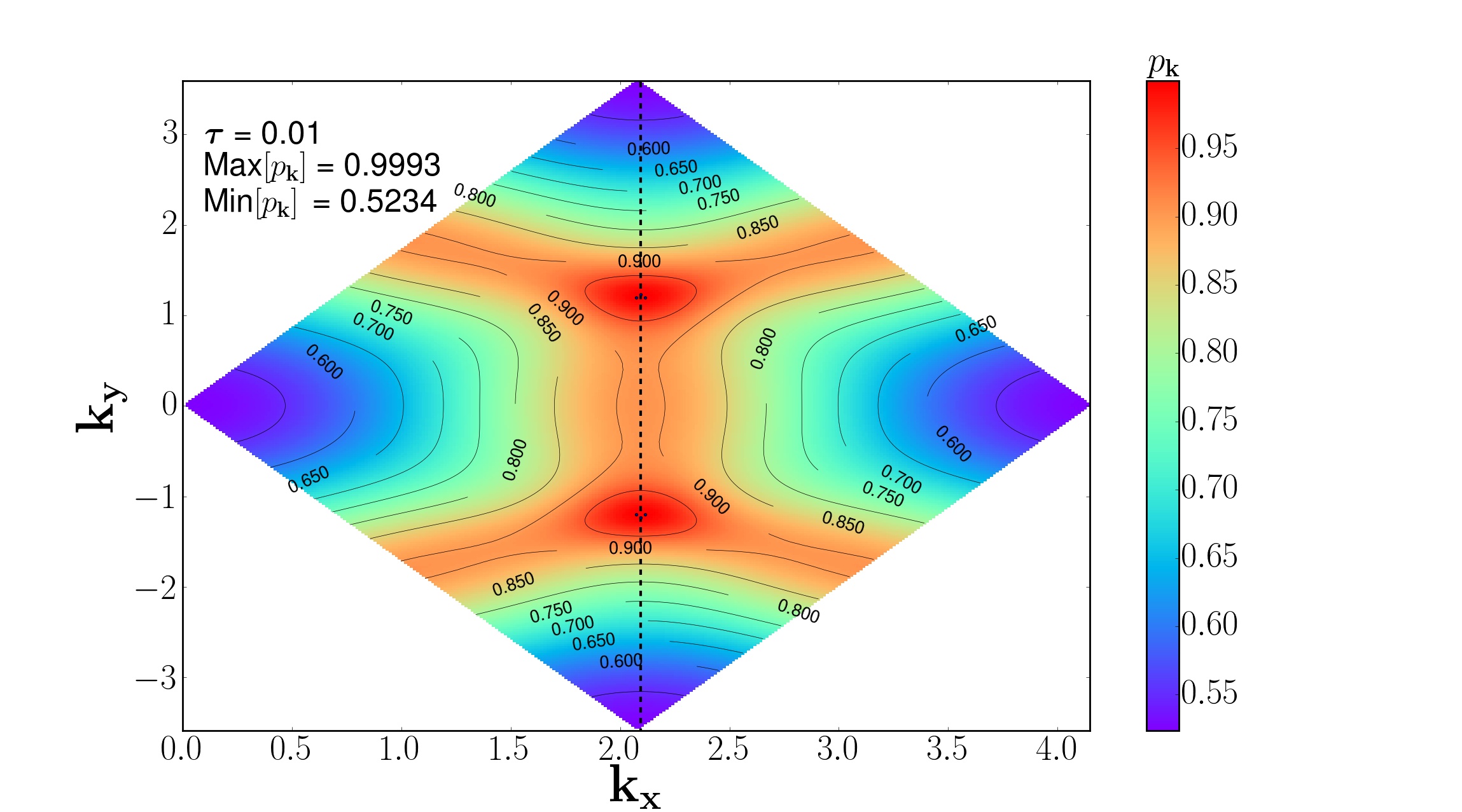}
\label{nodptpk}}\hfill
\subfigure[]{%
\includegraphics[width=0.49\textwidth,height=6.5cm]{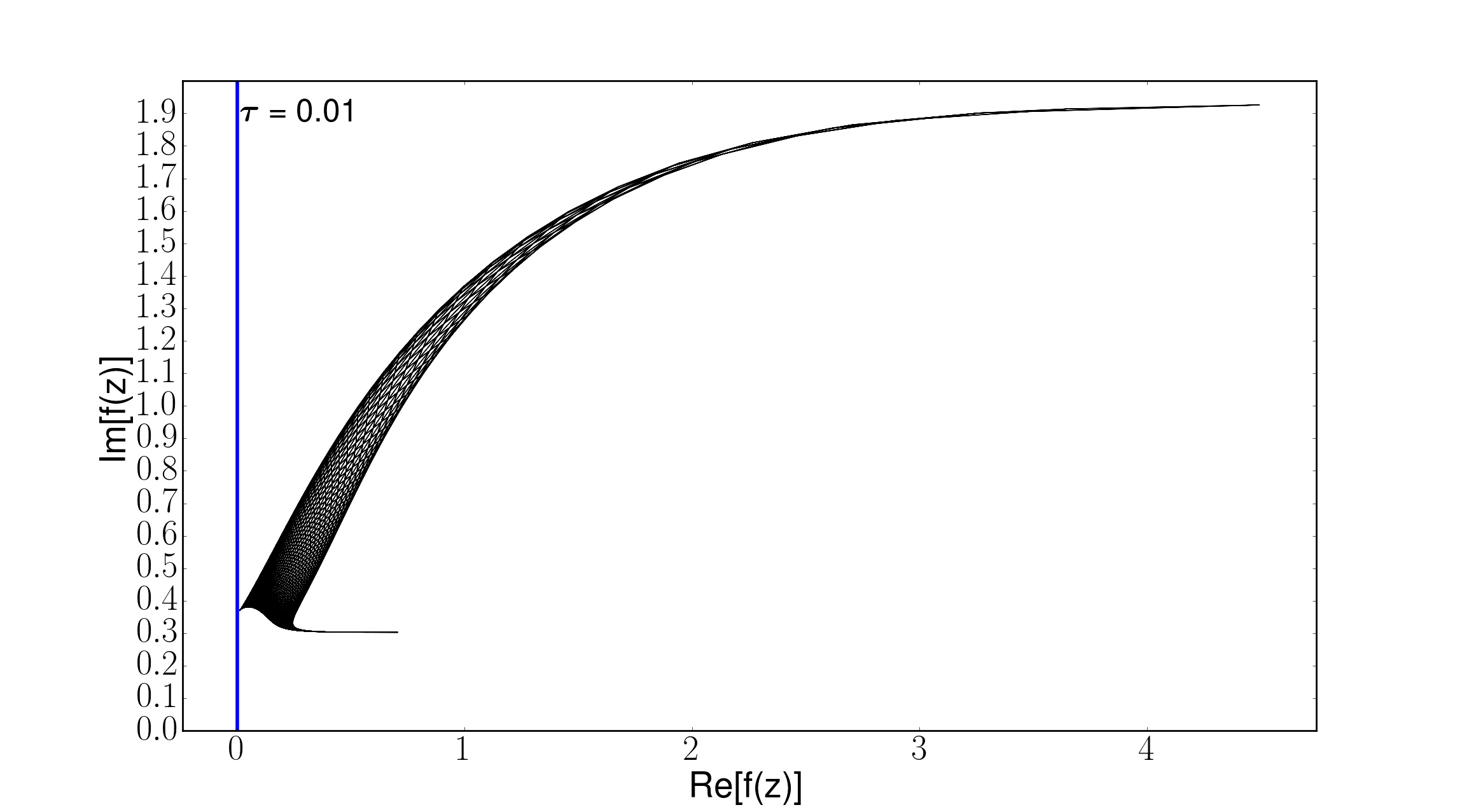}
\label{nodptfz}}
\begin{picture}(0,0)
\put(-154,25){\includegraphics[width=4.25cm,height=3.9cm]{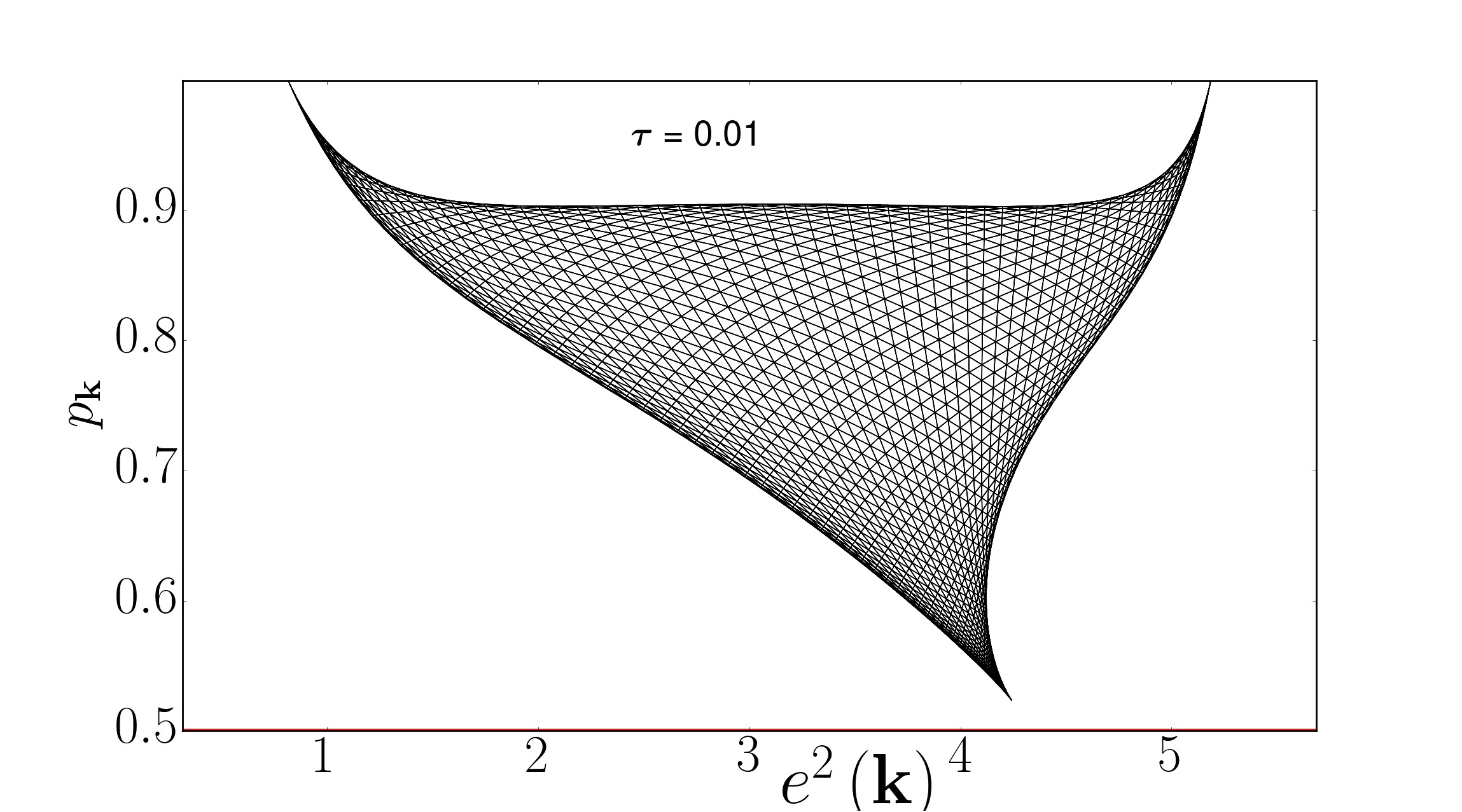}}
\end{picture}
\label{fig_nodptep}
\label{fig.nodpt}
\centering
\subfigure[]{%
\includegraphics[width=0.49\textwidth,height=6.5cm]{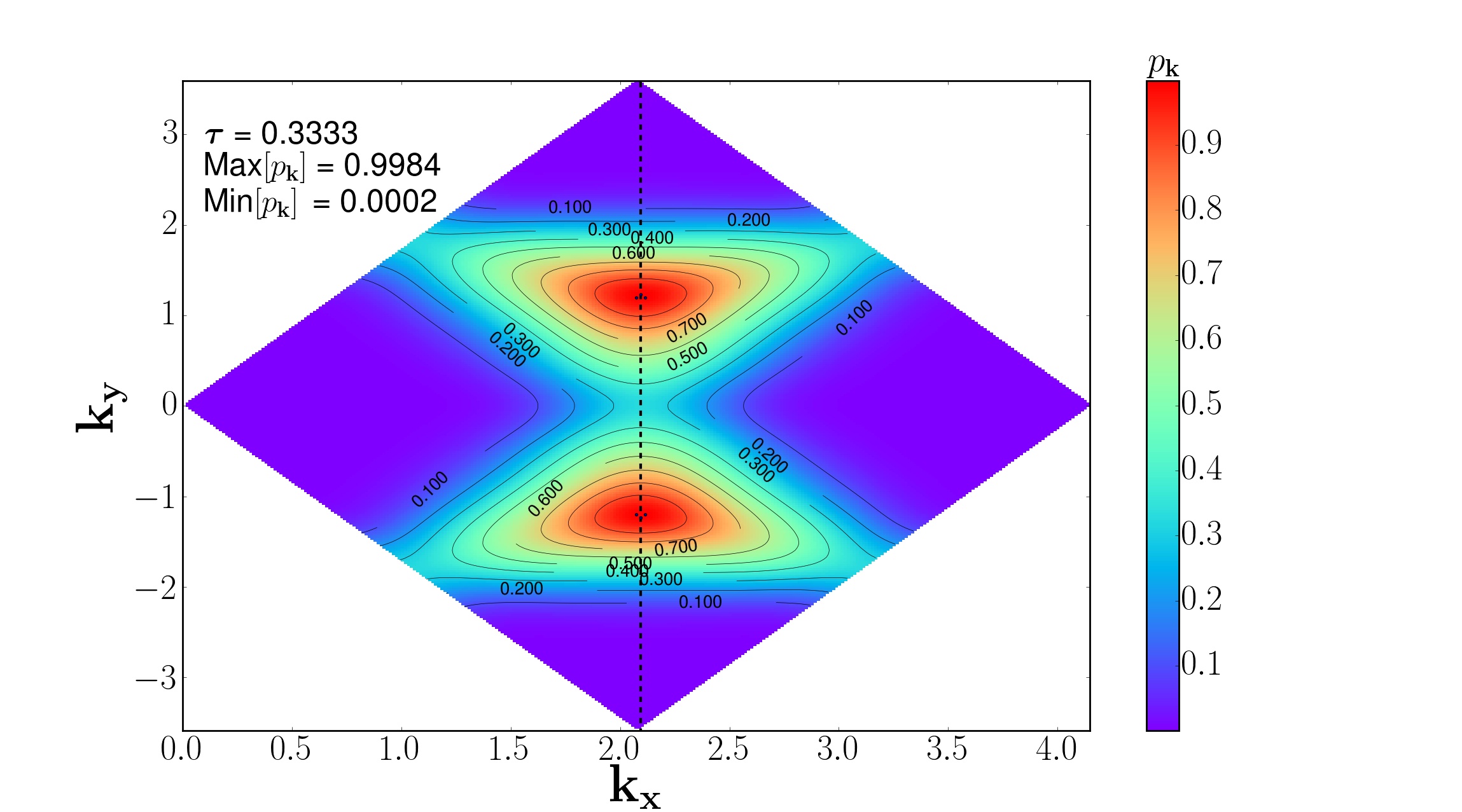}
\label{twodptpk}}\hfill
\subfigure[]{%
\includegraphics[width=0.49\textwidth,height=6.5cm]{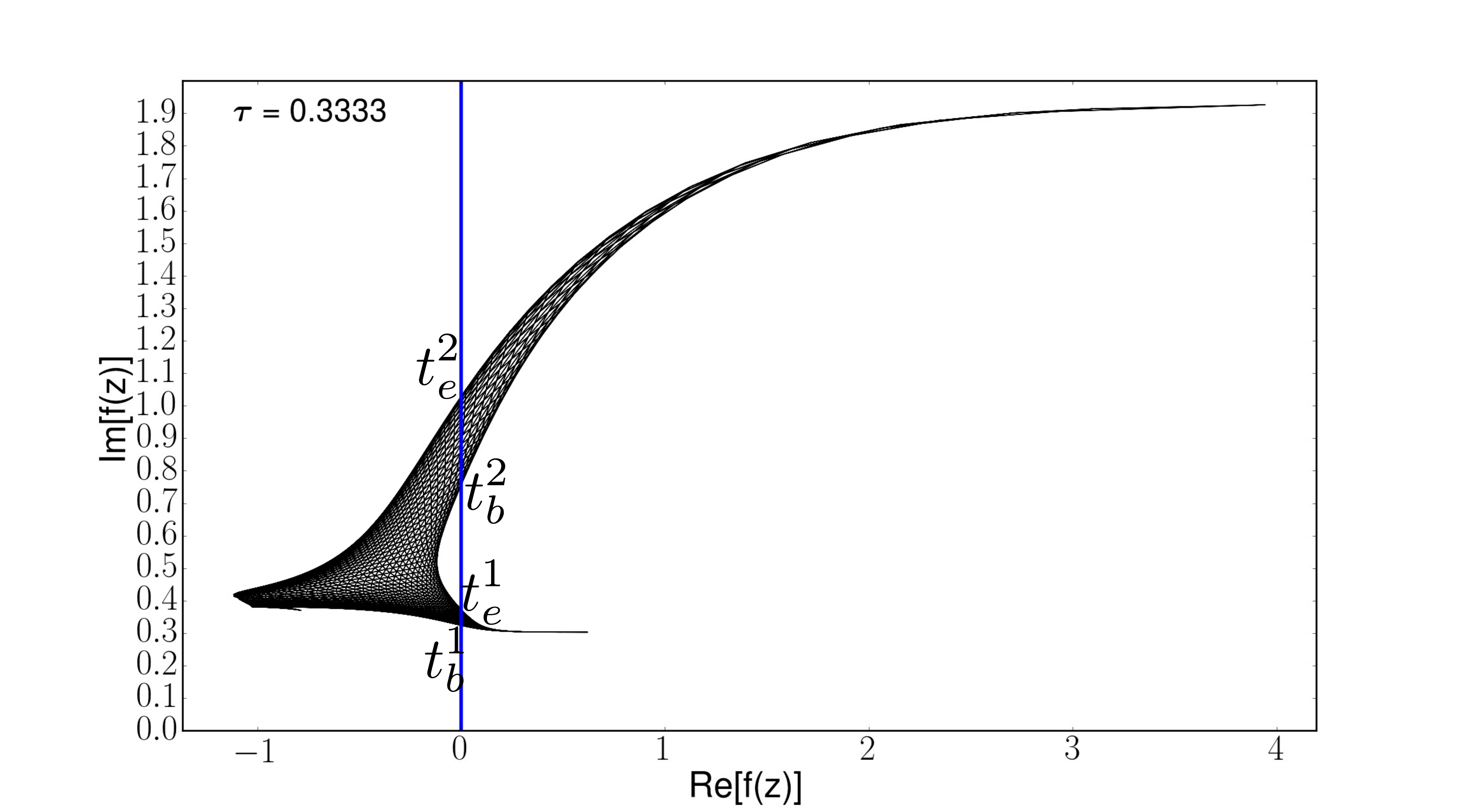}
\label{twodptfz}}
\begin{picture}(0,0)
\put(362,230){\includegraphics[width=4.2cm,height=3.65cm]{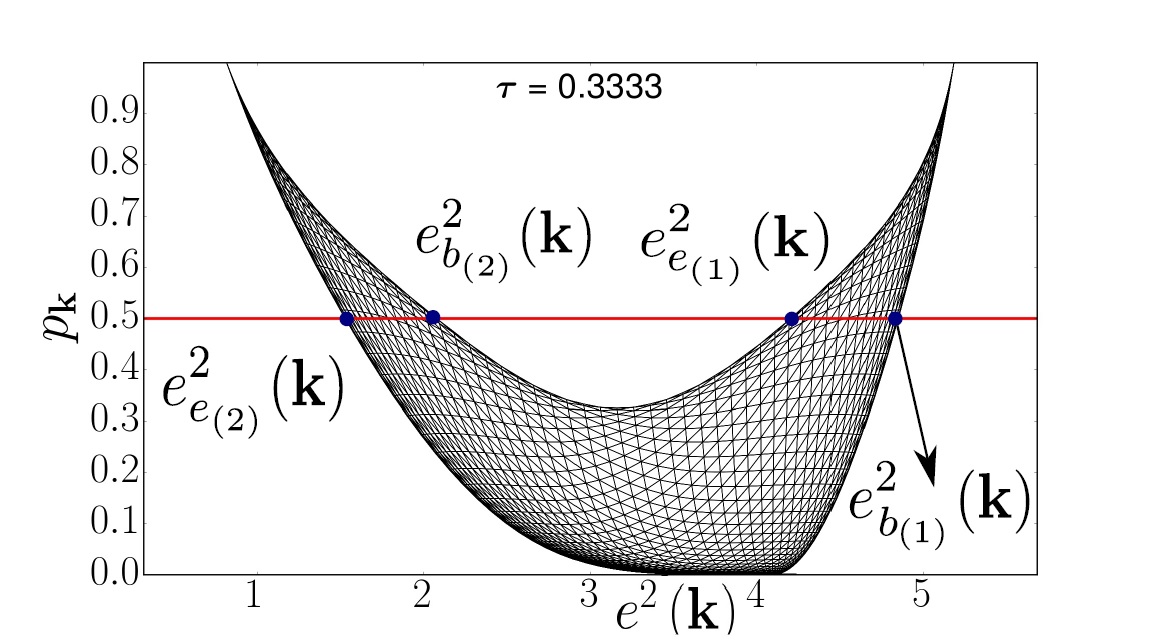}}
\end{picture}
\centering
\subfigure[]{%
\includegraphics[width=0.48\textwidth,height=6.5cm]{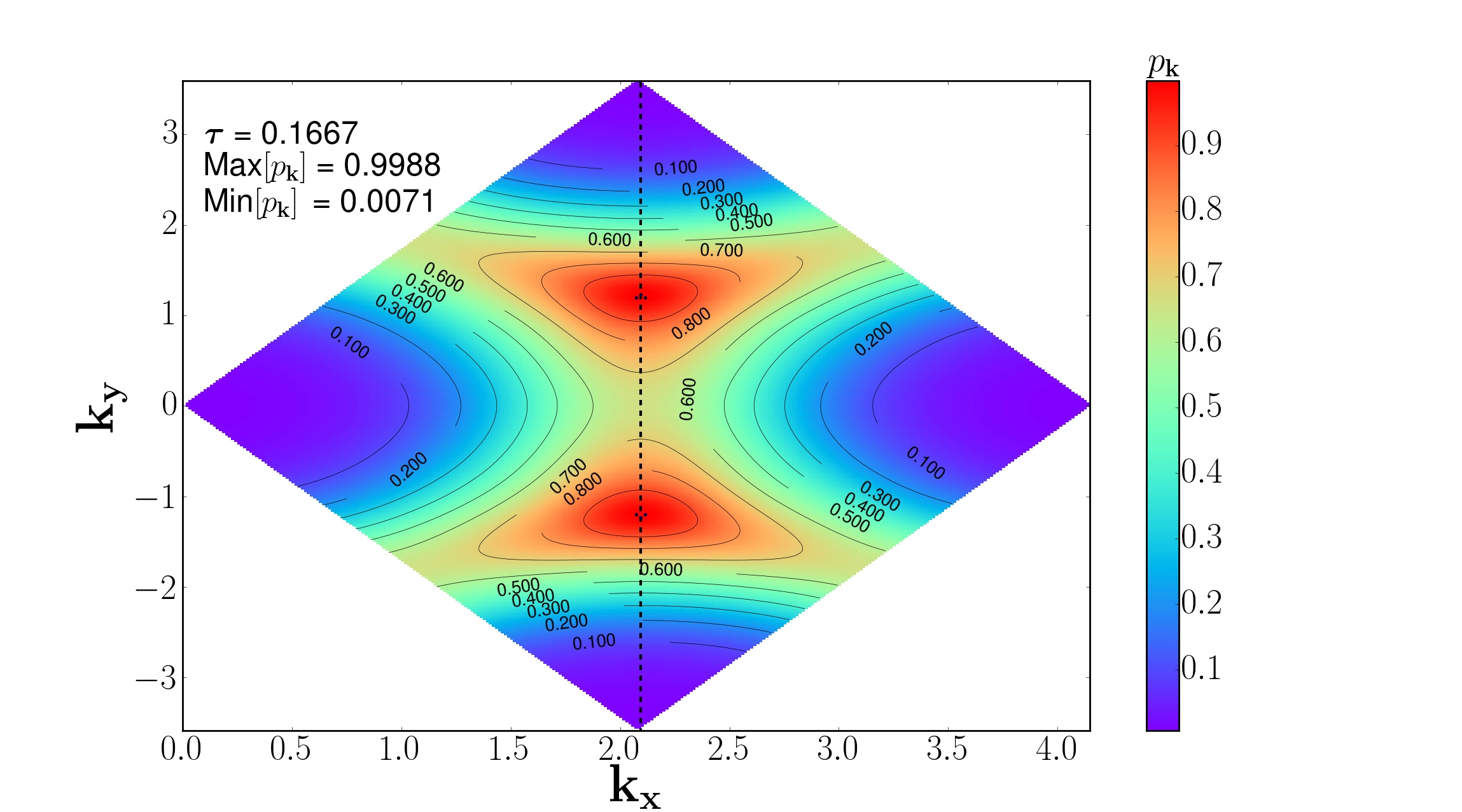}
\label{onedptpk}}\hfill
\subfigure[]{%
\includegraphics[width=0.48\textwidth,height=6.5cm]{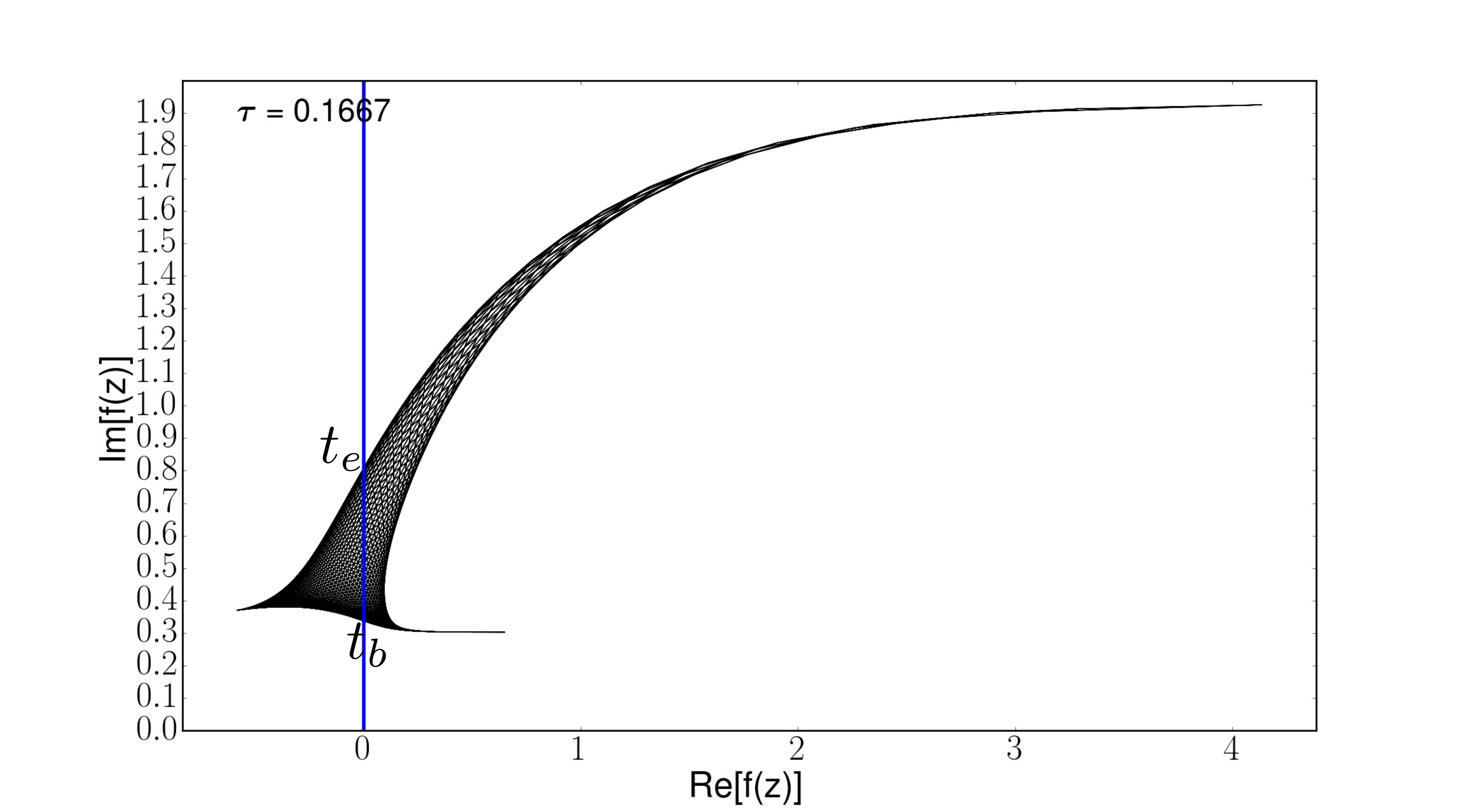}
\label{onedptfz}}
\begin{picture}(0,0)
\put(-153,25){\includegraphics[width=4.2cm,height=3.65cm]{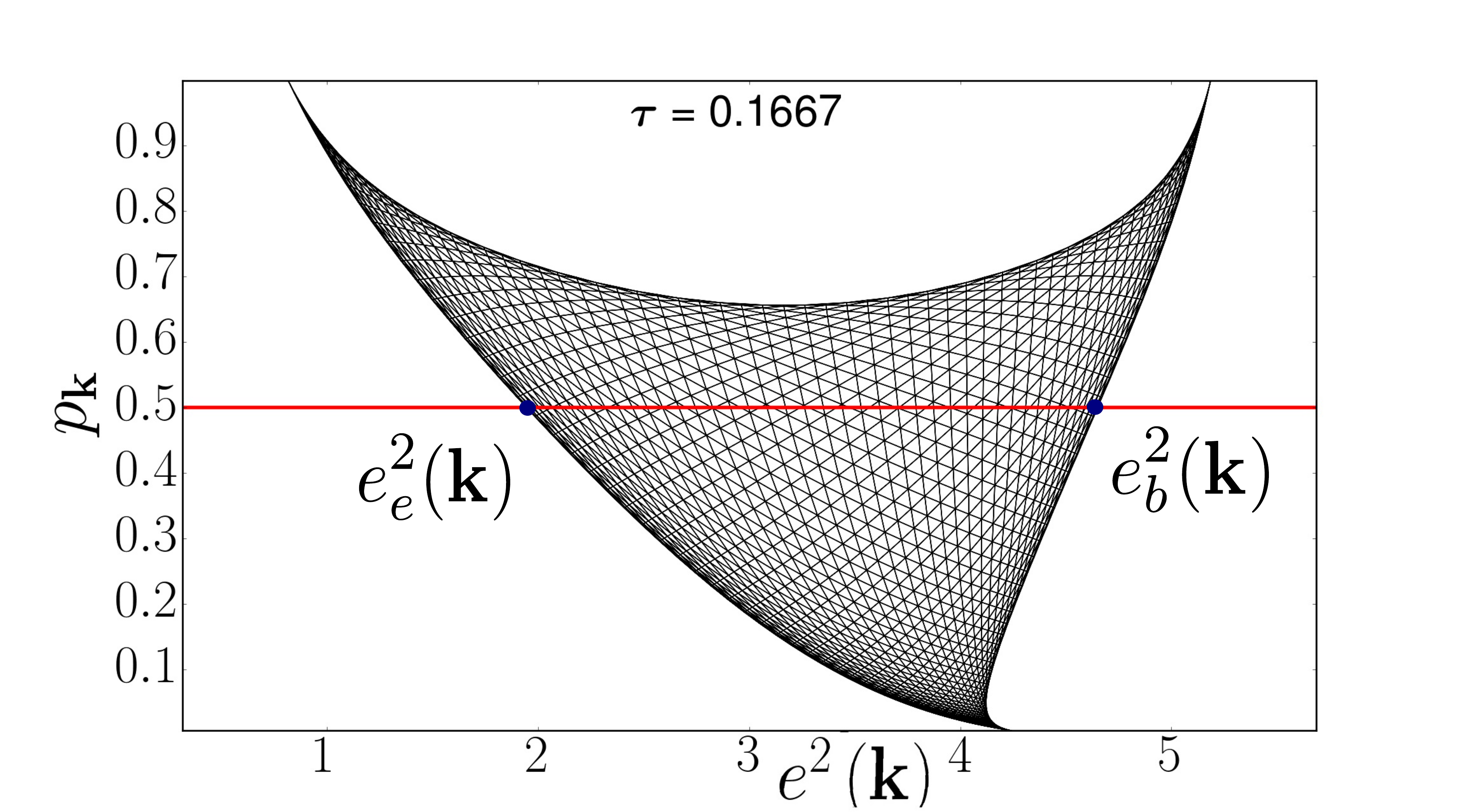}}
\end{picture}
\caption{~(color online) This set of figures presents the variation of $p_{\mathbf{k}}$ in the $k_x-k_y$ plane and no-DQPT,  re-entrant and internediate behaviours of FZs
for the sector  $n=0$ with $M_i=3, M_f=-3, \phi=1$ and   $L=100$. In Figs.~\ref{nodptpk} and \ref{nodptfz}, $\tau$ is   small ($<\tau_c^1$) so that the minimum value of $p_{\mathbf{k}}\geq 1/2$ for all values of $\mathbf{k}$ and  the area never crosses the real axis. The inset of Fig.~\ref{nodptfz}, validates the obseravtion that
there is no DQPT by showing that 
the area formed by $e^2(\mathbf{k})$ does not touch  the line $p_{\mathbf {k}}=p_{\mathbf {k^*}}=1/2$ for any value of the $\mathbf{k}$. On the contrary, Fig.
~\ref{twodptpk} shows that  in the extreme slow
limit, i.e., $\tau >\tau_c^2$,  $p_{\mathbf{k}}$ becomes less than 1/2 along the dotted line connecting two DPs and there is a re-entrant behavior of FZs evident in Fig.~\ref{twodptfz} : the
inset  shows four boundary points  of  $e^2({\mathbf{k}})$ on the line $p_{\mathbf {k^*}}=1/2$  leading to four  instants of time  $t_b^1$, $t_e^1$, $t_b^2$ and $t_e^2$, (as obained from Eq.~\eqref{eq_time}) where Re$[f'(t')]$ is non-analytic (see Fig.~\ref{fig_nonanal1}). For  $\tau_1^c < \tau <\tau_c^2$, Fig.~\ref{onedptpk} shows that
 $p_{\mathbf{k}}$ never becomes less that $1/2$ along the dotted line  and consequently, the area of FZs cross the real time axis only once (Fig.~\ref{onedptfz}): the inset shows the boundary points those lead to
$t_b$ and $t_e$, where Re$[f'(t')]$ is non-analytic as shown in Fig.~\ref{fig_nonanal2}.}
 \label{fig_results}

\end{figure*}

\begin{figure*}[]
\centering
\subfigure[]{%
\includegraphics[width=.45\textwidth,height=6cm]{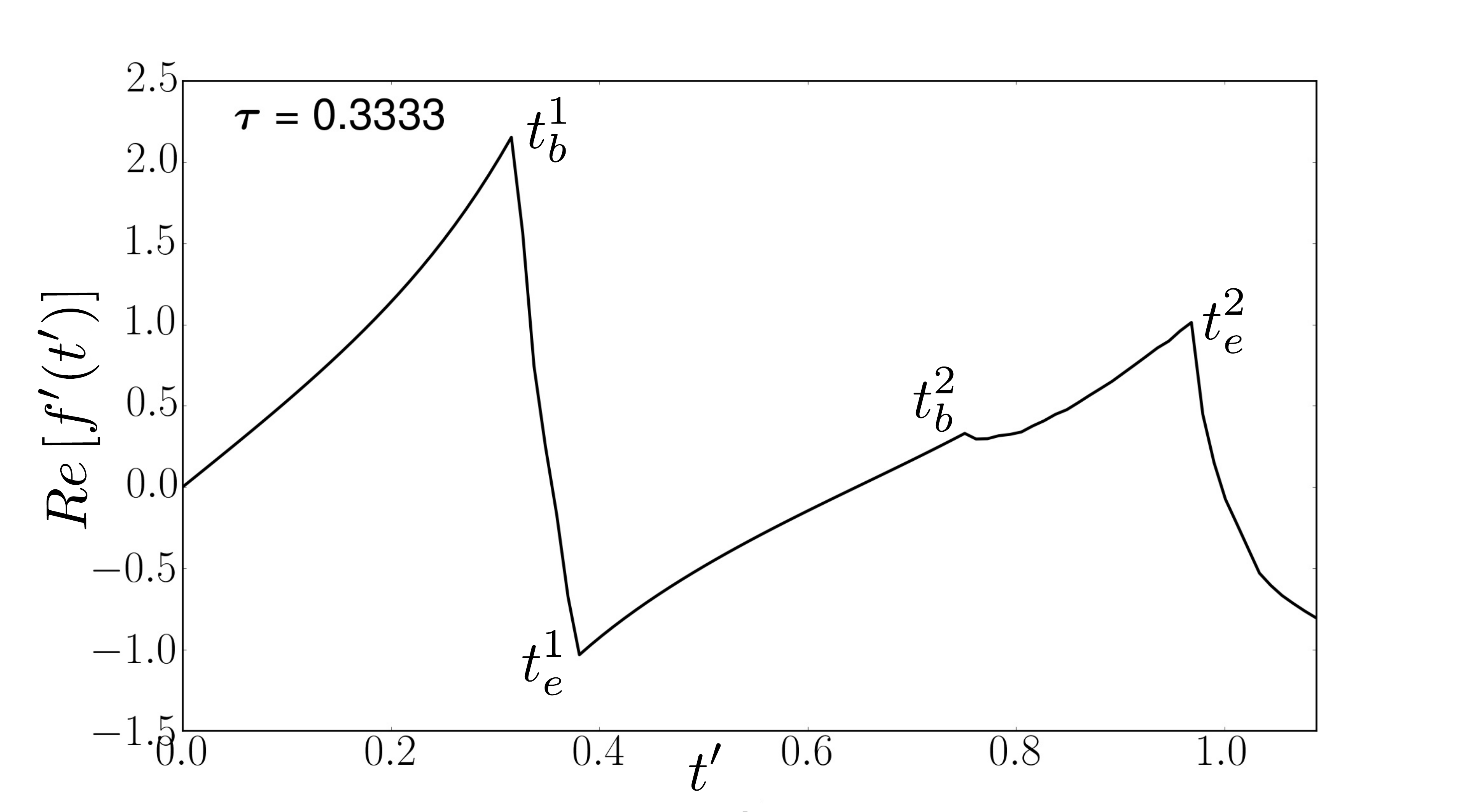}
\label{fig_nonanal1}}
\hfill
\quad
\subfigure[]{%
\includegraphics[width=.45\textwidth,height =6cm]{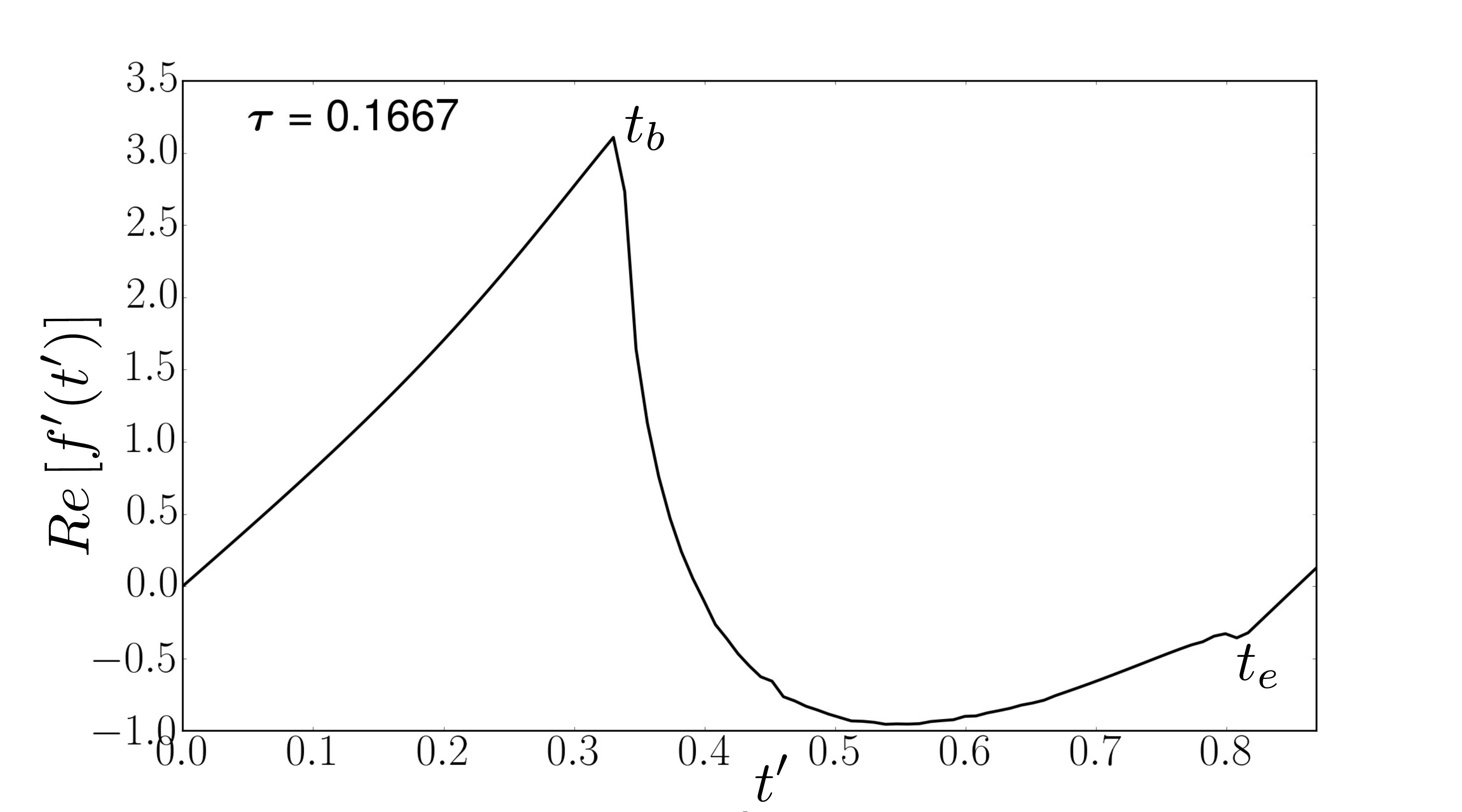}
\label{fig_nonanal2}}
\caption{The non-analyticities (cusp singularities)  in Re$[f'(t')]$ in the re-entrant and the intermediate  situations (as predicted in Figs.~\ref{twodptpk}-\ref{onedptfz}) are
shown in Figs. \ref{fig_nonanal1} and \ref{fig_nonanal2}, respectively, for $n=0$. Here, although we have used same $\tau$, $\phi$, $M_i$ and $M_f$ values as in
Fig.~\ref{fig_results}, $L$ is  chosen to be $1000$ to accentuate the instants of real time at which sharp non-analyticities appear. }
\end{figure*}

\section{Results and Analyses}

\label{sec_results}

 \begin{figure*}[]
\centering
\subfigure[]{%
\includegraphics[width=0.49\textwidth,height=7cm]{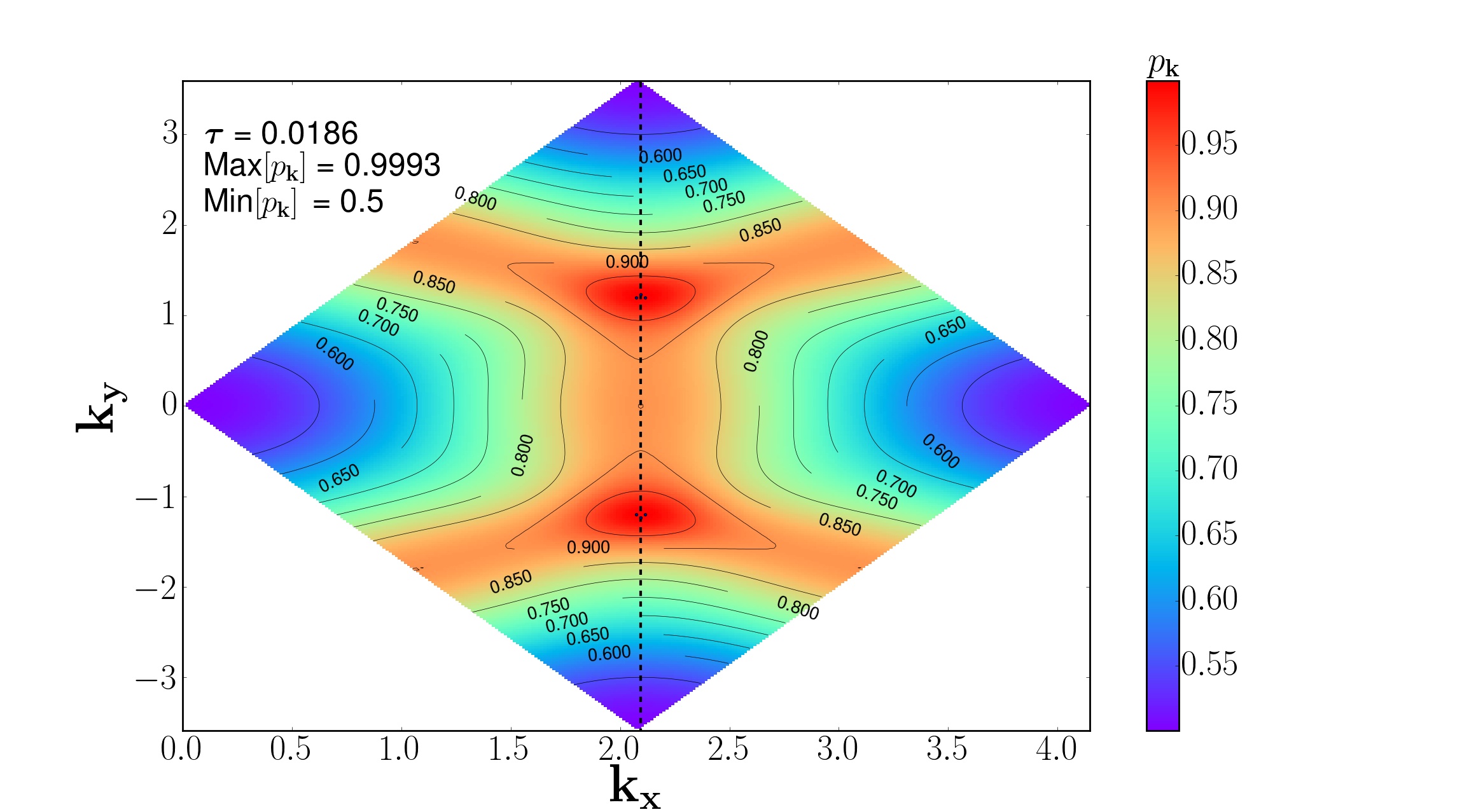}
\label{tau1cpk}}\hfill
\subfigure[]{%
\includegraphics[width=0.49\textwidth,height=7cm]{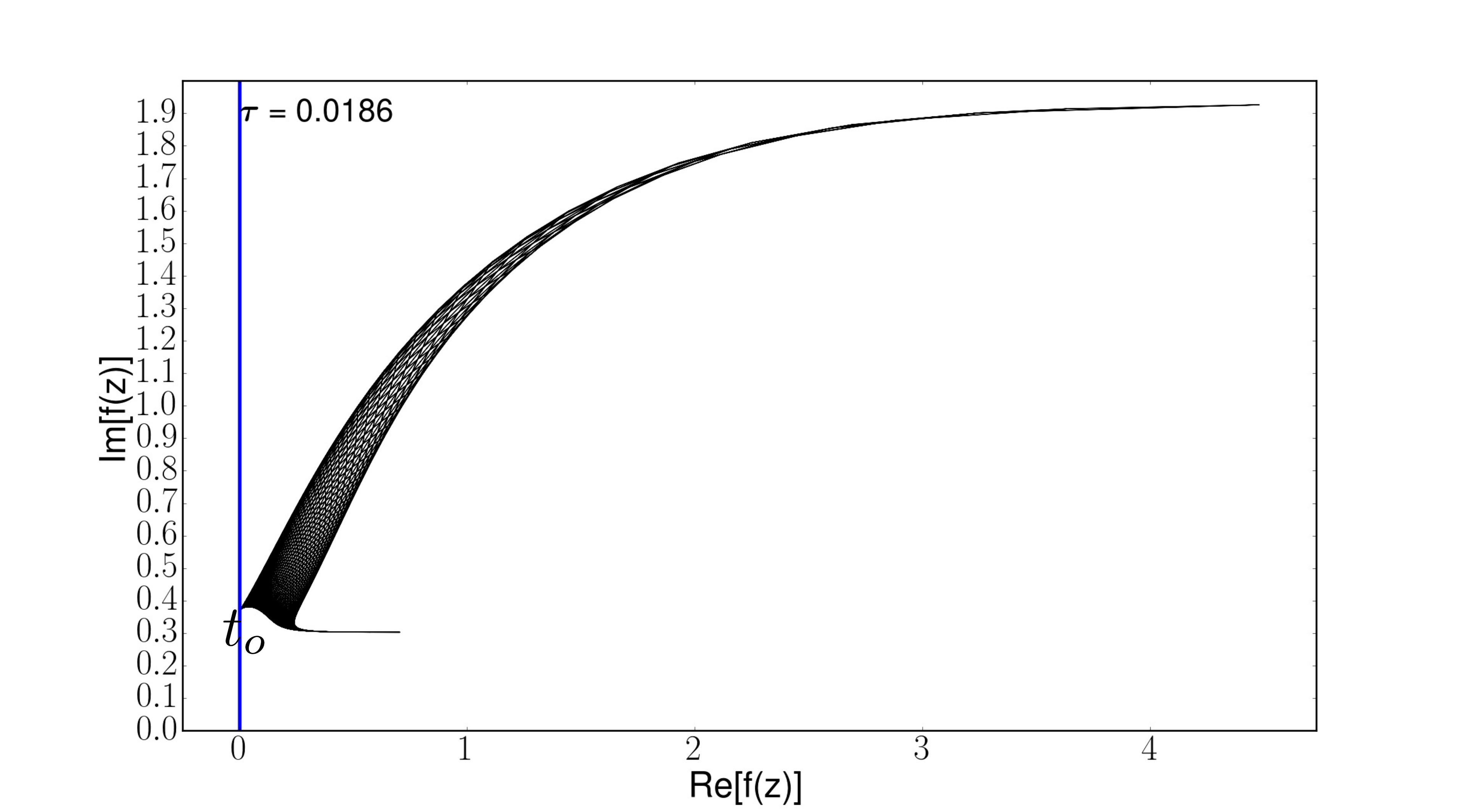}
\label{tau1cfz}}
\begin{picture}(0,0)
\put(-156,25){\includegraphics[width=4.25cm,height=4cm]{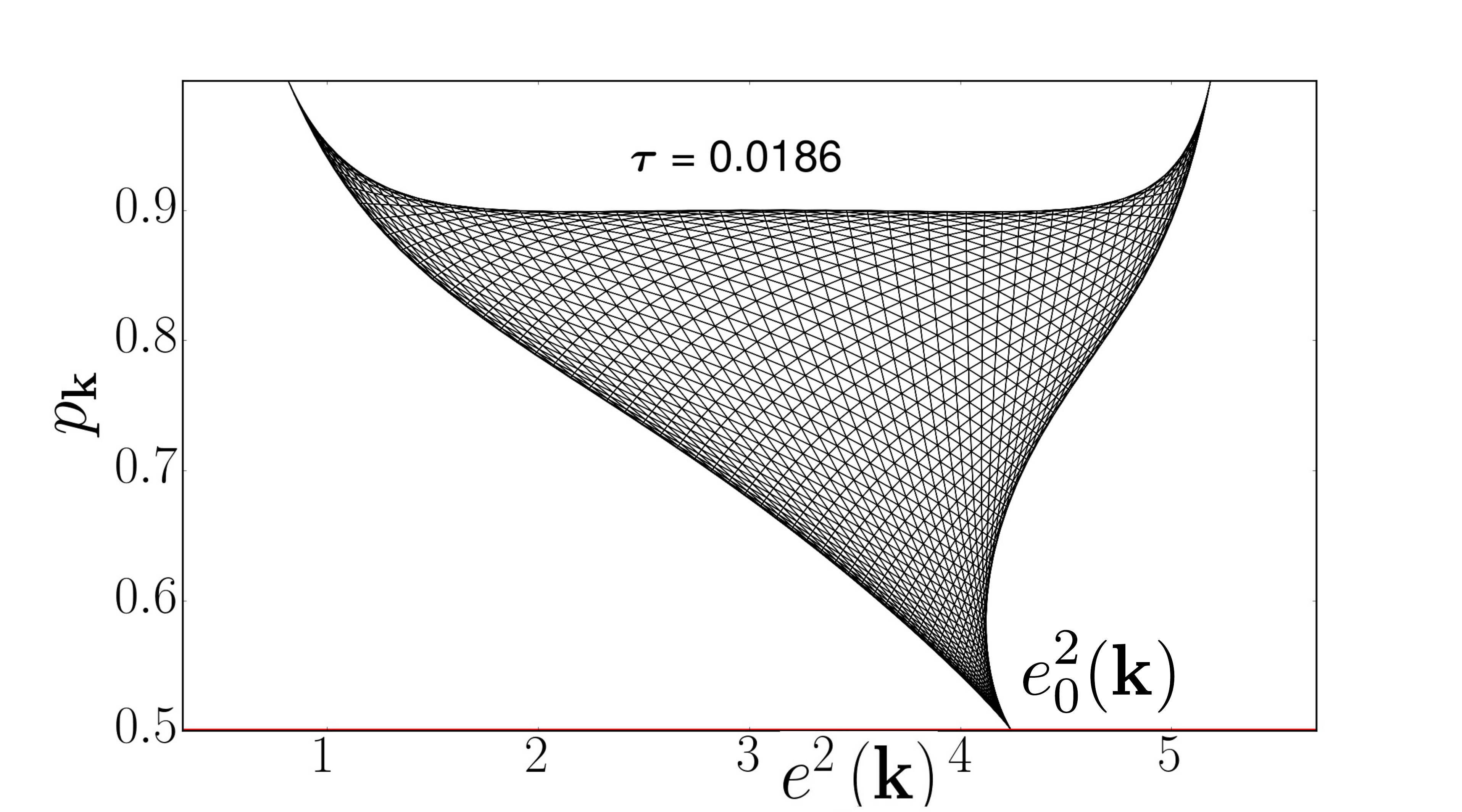}}
\end{picture}
\label{fig_nodptep}
\label{fig.nodpt}
\centering
\subfigure[]{%
\includegraphics[width=0.49\textwidth,height=7cm]{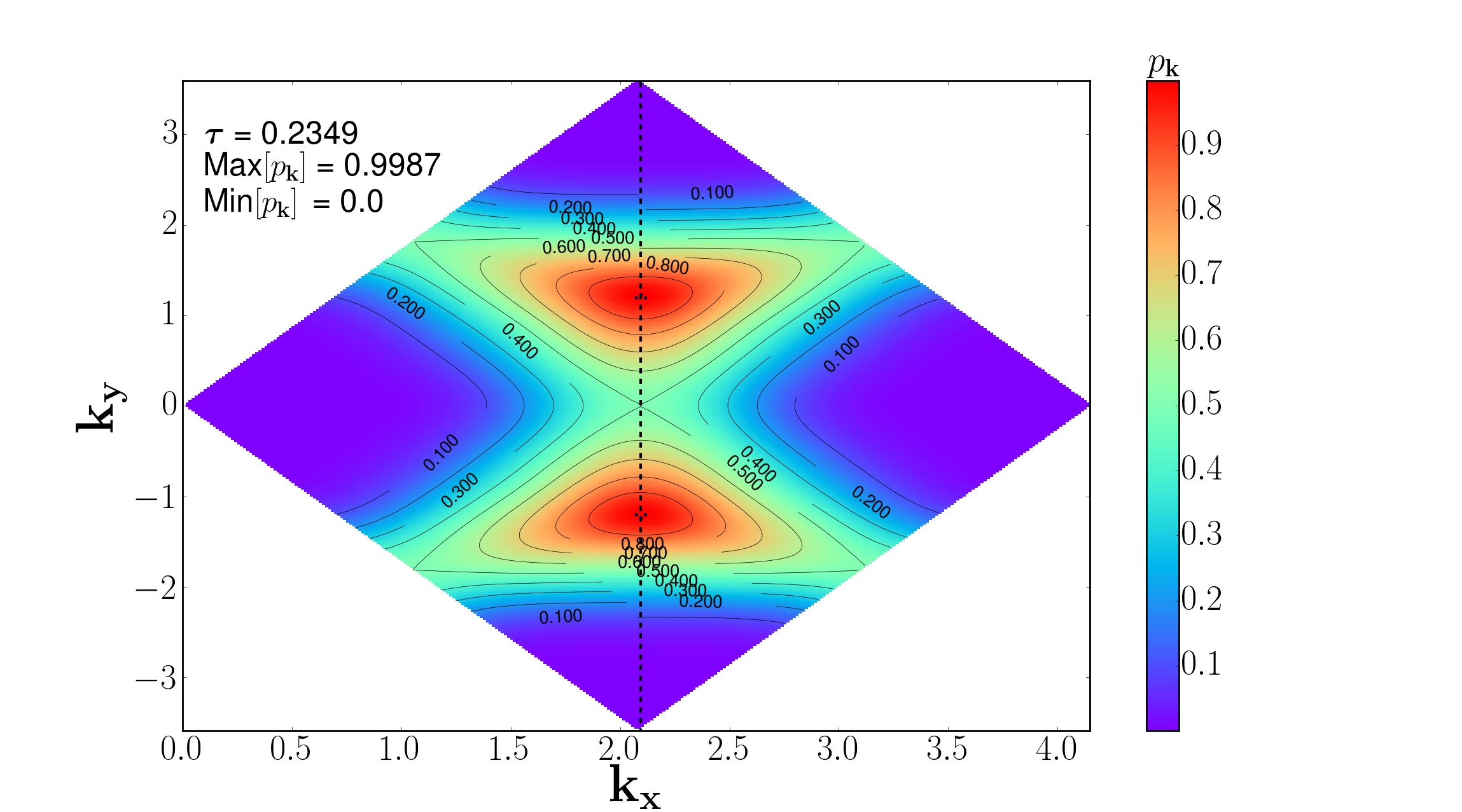}
\label{tauc2pk}}\hfill
\subfigure[]{%
\includegraphics[width=0.49\textwidth,height=7cm]{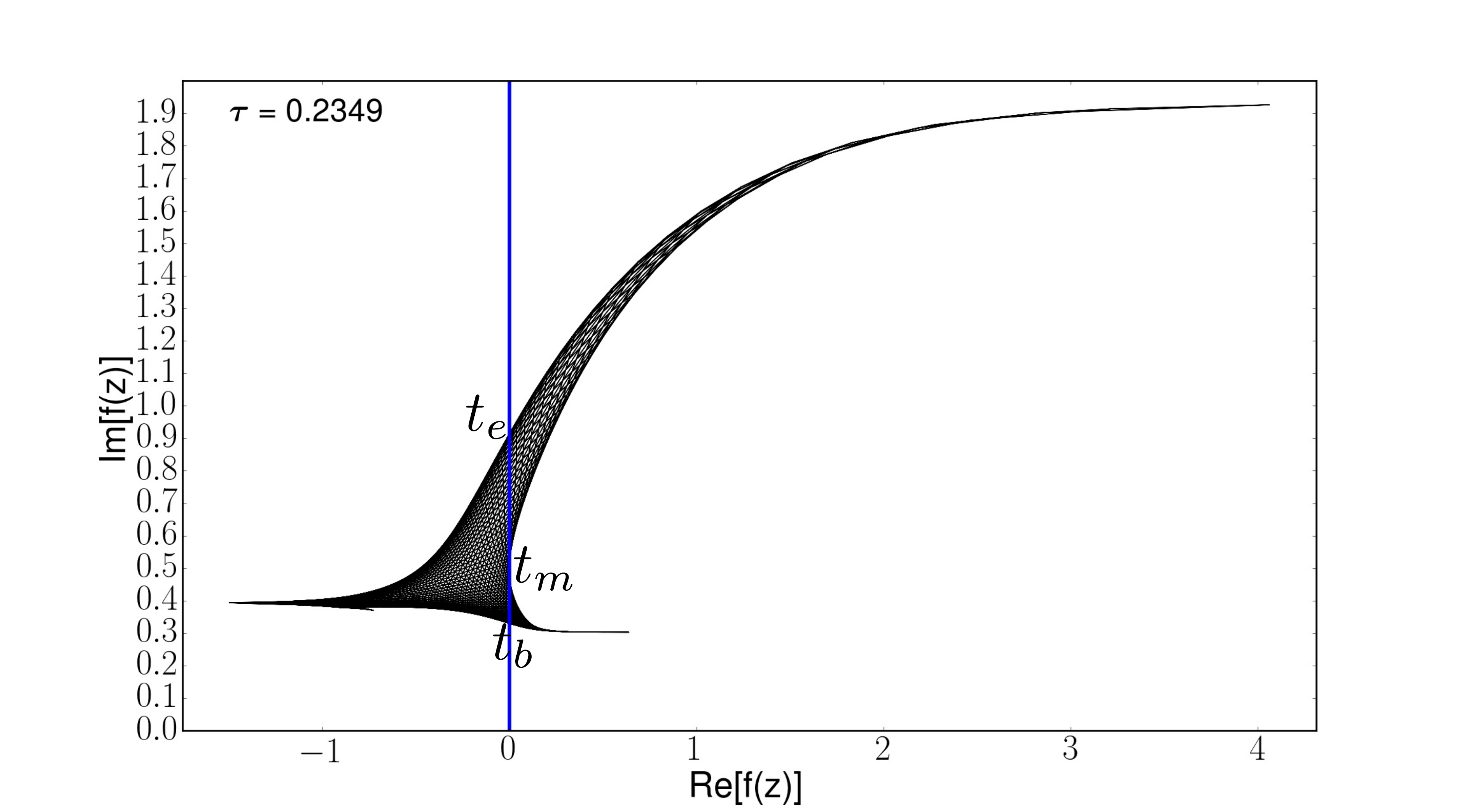}
\label{tauc2fz}}
\begin{picture}(0,0)
\put(114,40){\includegraphics[width=4cm,height=3.8cm]{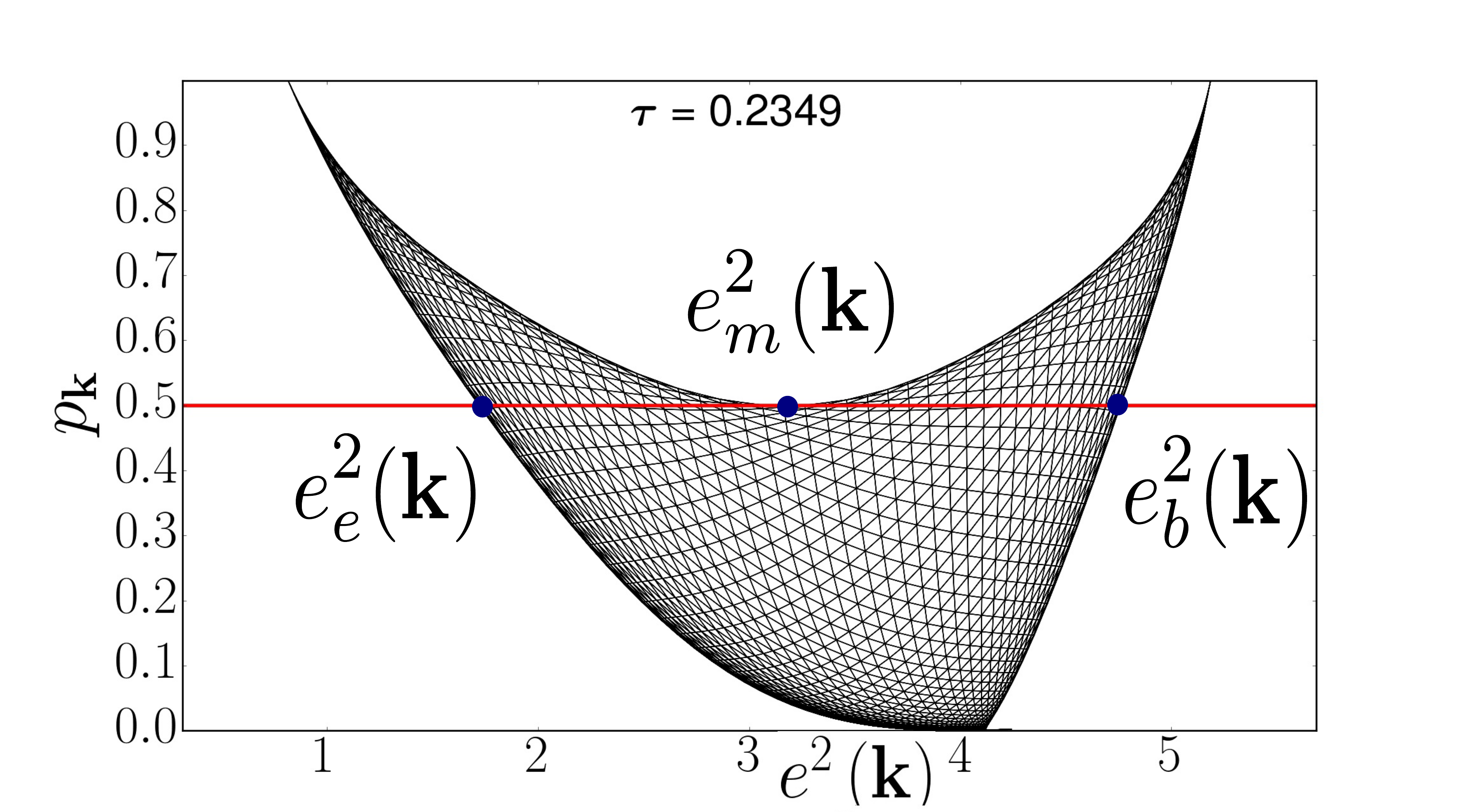}}
\end{picture}
\label{fig_twodptep}
\label{twodpt}
\caption{~(color online) This set of figures presents the scenario occurring at the critical values of $\tau$, namely, $\tau_c^1$ and $\tau_c^2$, with $M_i=3, M_f=-3, \phi=1$ and   $L=100$. As shown in Fig.~\ref{tau1cpk},  at
$\tau=\tau_c^1$,  the excitation probability $p_{\mathbf{k}}$ becomes $1/2$ for the mode $k_x=k_y=0$ (the TRIM point at the corner of BZ in Fig.~\ref{fig_BZ}) for which $M_H=0$.  Fig.~\ref{tau1cfz} shows that the area of FZs touch the imaginary (real time) axis for this mode  at the instant $t_m$; the 
inset shows  that $e^2(\mathbf{k})$ touches the $p_{\mathbf{k^*}} =1/2$ line for this momentum mode. On the contrary, at $\tau=\tau_c^2$ (Fig.~\ref{tauc2pk}), $p_{\mathbf{k}}$ just
becomes 1/2 
for the TRIM point at the centre of the BZ.  Consequently, as shown in (Fig.~\ref{tauc2fz}), the internal boundary
of the area of FZs touches the real time axis for this mode; this is more transparently depicted in the inset, there is an additional boundary point for the energy
$e^2_m(\mathbf{k})$ (that corresponds to the instant $t_m$) when compared to the inset of the Fig.~\ref{onedptfz}.}
\end{figure*}

In this section we present the main results of our paper   {based on the quenching scheme $M(t)=-t/\tau$ with $M_i=3$  and $M_f=-3$ with $\phi=1$ as shown
in Fig.~\ref{fig_haldane_phase}} and discuss three  cases which illustrate how the appearance of FZs vary with the inverse rate of quenching  $\tau$ as we tune it  from the sudden limit (when $\tau\rightarrow 0$) to the extreme slow limit (when $\tau\rightarrow\infty$) ramping
the system linearly  across two massless DPs (see Fig.~\ref{fig_results}). 

As is evident from the figures in all situations $p_\mathbf{k}$ is $1$ at the DPs and smaller than $1$ for modes away from them.  {Consequently, from Eq.~\eqref{eq_fisher_zero1}, we conclude that  FZs corresponding to DPs would tend to $+\infty$, while for the modes with  $p_{\mathbf{k}} \to 0$ the FZs
would approach $-\infty$, yielding the possibility of FZs extending from $-\infty$ to $+\infty$ in the thermodynamic limit.
We note that the 
maximum value of $p_{\mathbf{k}}$ is of the order of $.999$ (i.e., nearly unity) near the DPs ensuring  that our numerical scheme is significantly accurate.

{Let us first  consider the  extreme limit, i.e, the sudden quenching limit where the value of $\tau \to 0$. Therefore, although the system is quenched across
both the QCPs, as shown in Fig.~\ref{nodptpk},
the rapid rate of quenching never allows the minimum value of $p_{\mathbf{k}^*}$ to become less than 1/2; consequently,  area of FZs never cross
the real time axis (as shown in Fig. \ref{nodptfz}) and  hence there is no non-analyticity either in Re$[f(t)]$ or in its time derivative resulting in a complete absence of DQPTs.}

Now considering the other extreme limit, i.e., the slow limit (see Fig. \ref{twodptpk}), let us follow the contour plot of the excitation probability, $p_\mathbf{k}$ along the $k_x=\frac{2\pi}{3}$ line. Every time the system goes from a value of $p_\mathbf{k} = 0$ to $p_\mathbf{k} = 1$, the excitation probability attains the value of $p_\mathbf{k} =p_\mathbf{k^*} = {1}/{2}$. These values of $k^*_x$ and $k^*_y$ form two lobes (closed contours) of $p_\mathbf{k^*} = {1}/{2}$ centred around the two  DPs.  Referring to Eq.~\eqref{eq_fisher_zero} and the discussion following it, one can explain the observation, as shown in Fig. \ref{twodptfz},  that the area of FZs (for $n=0$) crosses the real time axis twice 
which means that their boundaries cross four times;   this behavior of the areas of FZs is referred to as re-entrances.
 Referring to the inset of Fig.~\ref{twodptfz}, the line $p_\mathbf{k^*}=\frac{1}{2}$  cuts the area generated by $e^2(\mathbf{k})$ multiple times leading to  a continuum band or range of $\mathbf{k^*}$ with four boundary values of $e^2(\mathbf{k^*})$; 
  these boundary values  when substituted in Eq.~\eqref{eq_time} (with $n=0$)
   lead to four time scales $t_b^1$, $t_e^1$, $t_b^2$ and $t_e^2$, corresponding to upper and lower lobes of  $p_\mathbf{k^*} = {1}/{2}$ contours. Consequently, Re$[f'(t)]$ shows cusp singularities
 at these instants of time (see Fig. \ref{fig_nonanal1}).  Again, from Eq. \eqref{eq_time}, we immediately conclude that $t_b^1$ and $t_b^2$ corresponds to the maximum of the highest and the lowest energy bands, while  $t_e^1$  and $t_e^2$  are given in terms of the minimum of these two energy bands.
 
Focussing on the $n=0$ sector, we have so far analysed two limits of $\tau$: extreme fast  ($\tau < \tau_1^c$), when there is no DPT  and the extreme slow  ($\tau > \tau_2^c)$ when there are re-entrances.
But the most intriguing situation arises at intermediate values of $\tau=\tau_i$, where $\tau_1^c<\tau_i<\tau_2^c$. Here, although the system passes through two critical points, the FZs (for $n=0$) collate to form an area in the complex $z$ plane whose boundaries cross the real time axis at two pairs of time instants only (see Fig.~\ref{onedptfz}), in stark contrast with re-entrance of FZs 
shown in Fig.~\ref{twodptfz}.
 Carefully analysing this scenario 
 from Fig. \ref{onedptpk}, we note that 
 the crucial difference with the re-entrance case happens to be the fact
that there exists only  a {\it single} $p_\mathbf{k}=p_\mathbf{k^*}=1/2$, curve encircling both the DPs (QCPs). Focusing on the variation of $p_{\mathbf{k}}=1/2$ with $k_y$ on the line  $k_x={2\pi}/{3}$, 
we conclude that $p_\mathbf{k} $ gradually becomes smaller than $1/2$ with distance from DPs on either side (of the DPs) but is definitely greater than ${1}/{2}$ between them. The upper critical value $\tau_2^c$ is that value of $\tau_i$ for which the $p_\mathbf{k}$ becomes ${1}/{2}$ between the two DPs on the $k_x={2\pi}/{3}$ line at the point $k_y=0$: for $\tau>\tau_2^c$, we see the formation of two $p_\mathbf{k}=p_\mathbf{k^*}={1}/{2}$ contours encircling the two DPs emerging  from the single contour for $\tau < \tau_c^2$. Therefore, 
in the intermediate regime, the value of $\tau_i$  is such that there is only one $p_\mathbf{k}= p_\mathbf{k^*}={1}/{2}$ lobe enclosing the two DPs
resulting  in the creation of a single area of FZs. Once again, recalling Eq. \eqref{eq_time}, the lower (upper) boundary of the area that crosses the real time axis is determined by the maximum (minimum) of $e^2(\mathbf{k})$ on the line $p_{\mathbf{k}}=1/2$ (inset of Fig.~ \ref{onedptfz}) and one observes cusp singularities
in Re$[f'(t)]$ only at two instants of real time as shown in Fig.~\ref{fig_nonanal2}.

\begin{figure*}[]
\centering
\subfigure[]{%
\includegraphics[width=.45\textwidth,height =6cm]{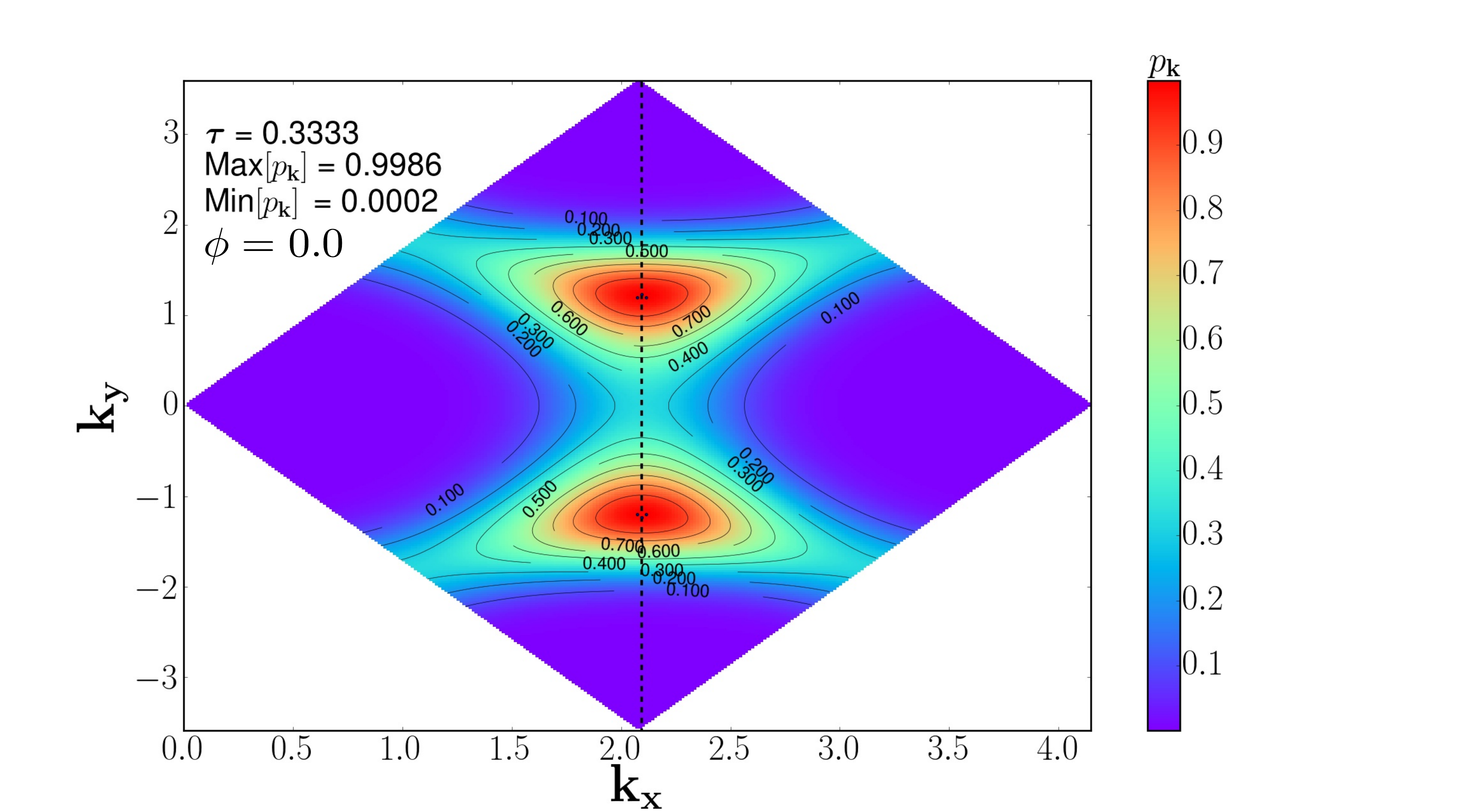}
\label{fig_phi01}}
\hfill
\quad
\subfigure[]{%
\includegraphics[width=.45\textwidth,height=6cm]{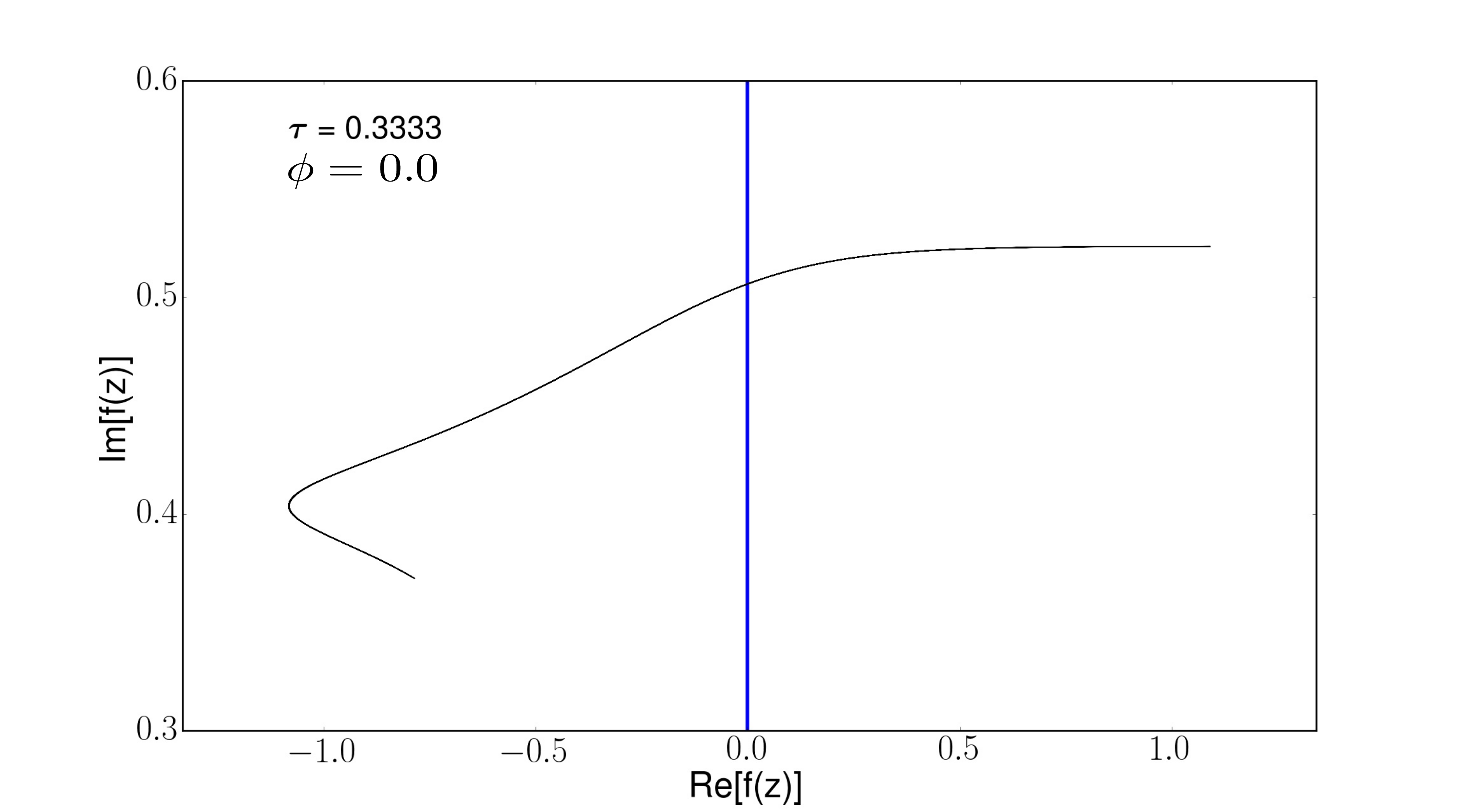}
\label{fig_phi02}}
\end{figure*}
\begin{figure*}
\centering
\subfigure[]{%
\includegraphics[width=.45\textwidth,height=6cm]{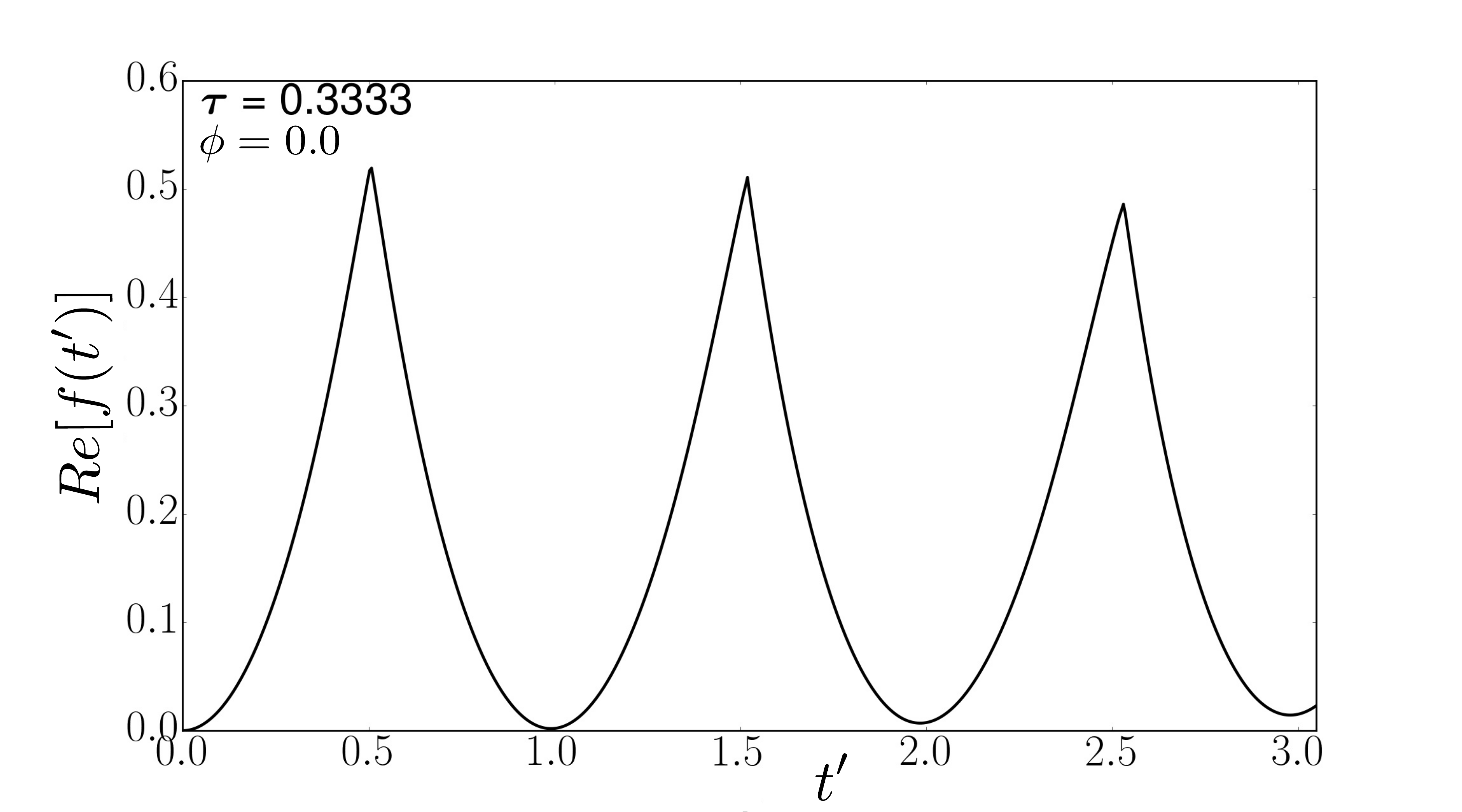}
\label{fig_phi03}}
\hfill
\quad
\subfigure[]{%
\includegraphics[width=.45\textwidth,height=6cm]{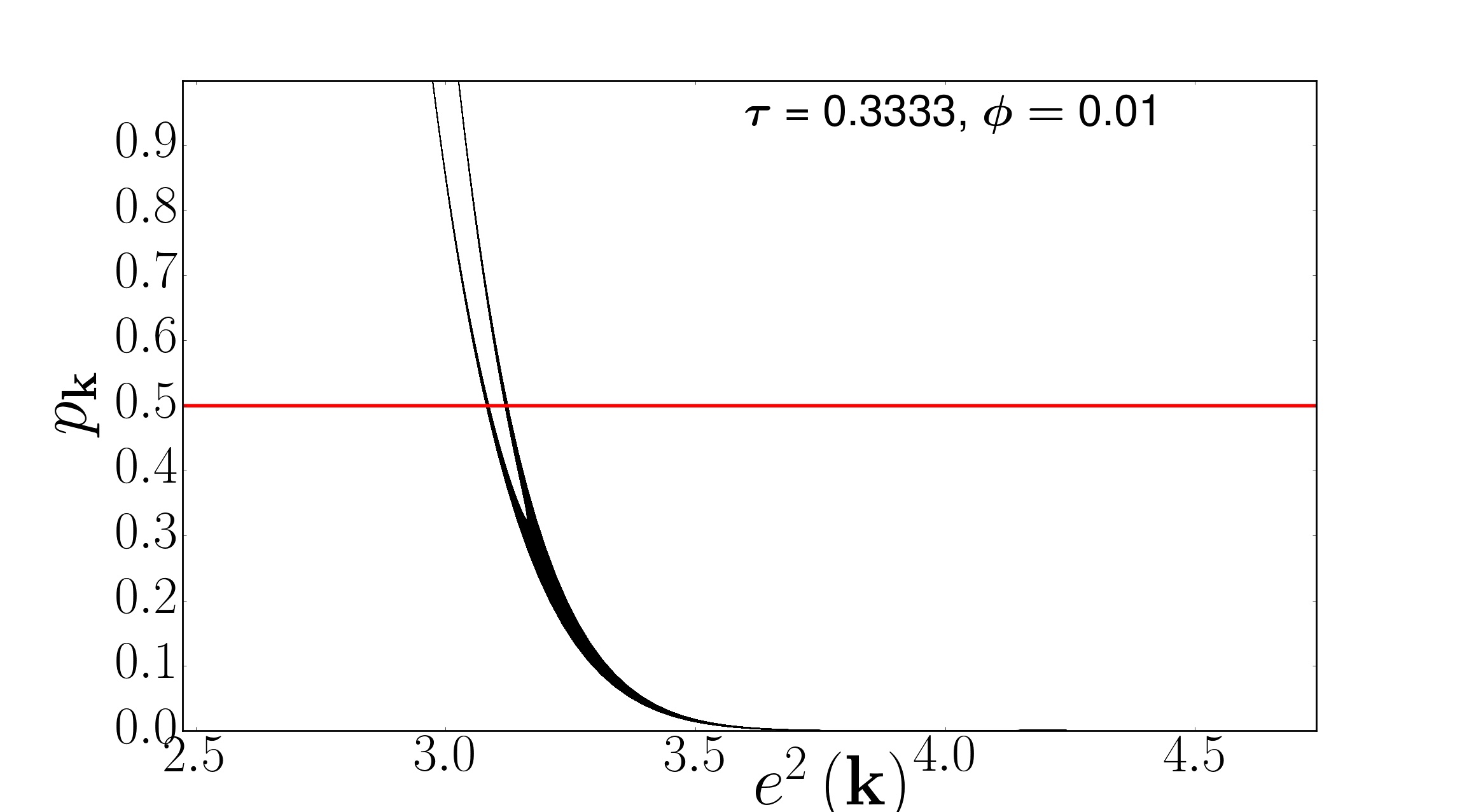}
\label{fig_phi04}}
\begin{picture}(0,0)
\put(-141.5,33){\includegraphics[width=3.9cm,height=3.7cm]{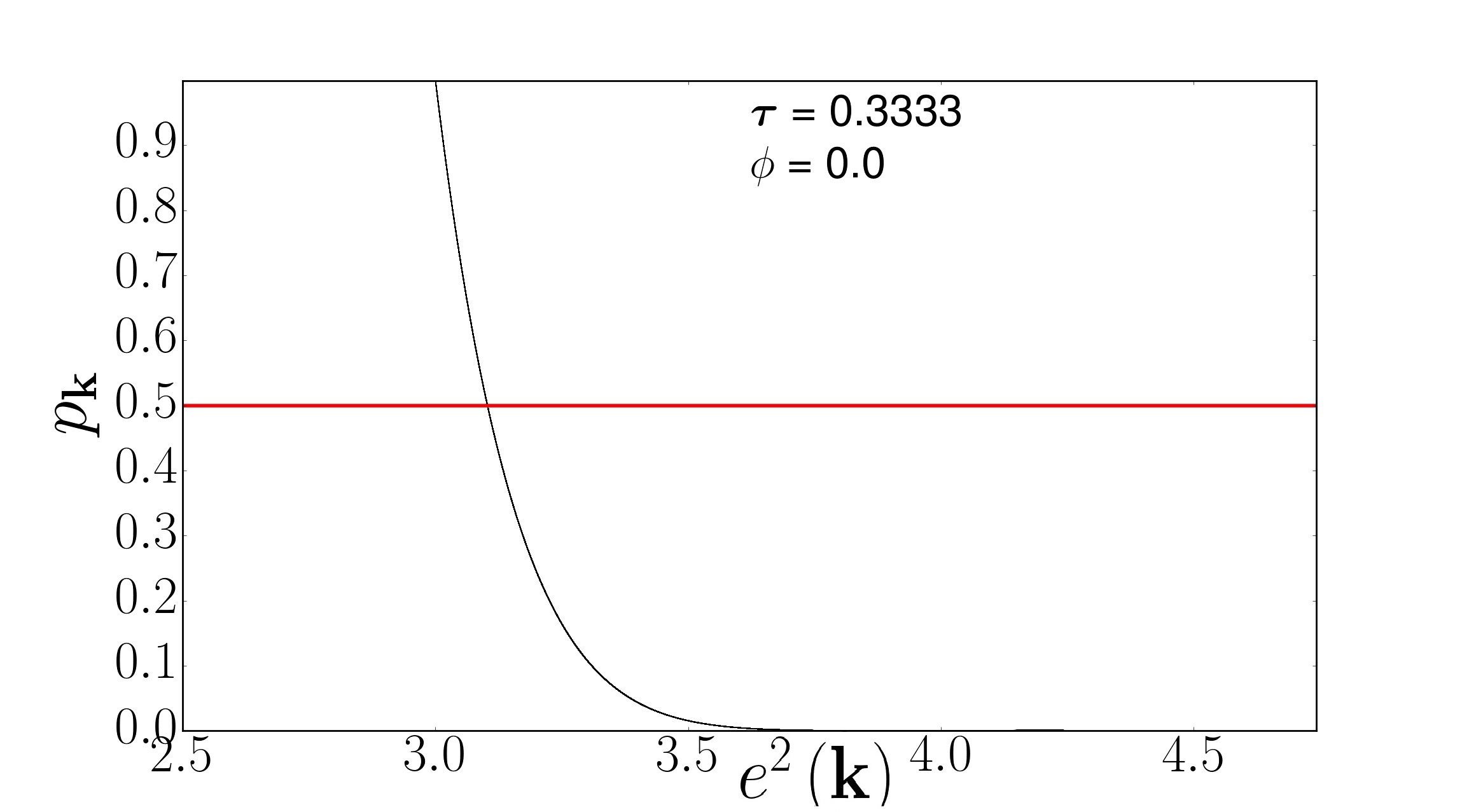}}
\end{picture}
\caption{~(color online) These figures demonstrate the effective 1D behavior that emerges when the parameter $\phi$ and consequently, $M_H$ vanishes; we choose
 $\tau=.3333$, so that for $\phi \neq 0$, there is a re-entrant behavior as presented Figs.~\ref{twodptpk} and \ref{twodptfz}. 
Fig.~\ref{fig_phi01} shows that  the excitation probability $p_{\mathbf{k}} \sim 1$, around both the DPs.  Nevertheless, as shown in Fig.~\ref{fig_phi02}, the FZs
constitute a line rather than an area; furthermore, the re-entrant behavior completely disappears. 
As elaborated in the text,  
in Fig.~
\ref{fig_phi03}), we show that   there exist sharp non-analyticities in Re$[f(t')]$ itself  at different instants of time and we choose $n=0,1$ and $2$.  In  Fig.~\ref{fig_phi04}, it is
illustrated that even for
very small $\phi$, there are four boundary points of $e^2(\mathbf{k})$ on the line $p_{\mathbf{k^*}}=1/2$, which shrink to one point as
soon as $\phi$ vanishes (inset). } 
\label{fig_zero_phi}

\end{figure*}

The question that naturally arises at this point is what determines the crossover values of the rate $\tau$ 
for which the transitions between the three different regions occur and
how do these scales depend on the parameter $\phi$ (or specifically $M_H$). Recalling the fact that the $p_{\mathbf{k}}$ is the smallest for the modes   {corresponding to the left corner TRIM point at 
$k_x=k_y=0$ fully included in the BZ   as shown in Fig.~\ref{fig_BZ}}, we focus on this TRIM point.  Since the minimum value of $p_{\mathbf{k}}$ must at least be $1/2$ for DQPTs to
appear, the condition Min$[p_{\mathbf{k}}]=1/2$, at this   {corner TRIM point} determines the scale $\tau_c^1$ as shown in Fig. \ref{tau1cpk}.  Min$[p_{\mathbf{k}}]$ 
at this TRIM, is independent of $M_H$, and  evidently
so is $\tau_c^1$.\\
On the other hand, the transition from the intermediate to the re-entrant phase taking place at $\tau_c^2$ occurs when the $p_\mathbf{k}={1}/{2}$ lobe enclosing the two DPs separate out at the other TRIM point  $k_x=2\pi/3, k_y=0$    {(located at the centre of  BZ)} with zero Haldane mass,  to form two distinct  lobes enclosing one DP each (Fig. \ref{tauc2pk}). 
When the value of $\tau$ is insufficient to make the $p_\mathbf{k}$ at this mode less than  $1/2$, the system remains in the intermediate phase, but at $\tau=\tau_c^2$, $p_{\mathbf{k}}$ becomes $1/2$ at this mode and the system enters the re-entrant phase for $\tau> \tau_c^2$. 
Hence, this mode being independent of $M_H$  again keeps $\tau_c^2$ invariant under the variation of $\phi$.  
We further emphasize that  both $\tau_c^1$ and $\tau_c^2$ are determined by TRIM points, {and 
 hence, are independent of the system size as long as it is ensured that the numerical scheme sharply includes these points while enumerating different modes of the BZ. Of course, these critical values do not get altered in the thermodynamic limit.}

Having established the independence of $\tau_c^1$ and $\tau_c^2$ on the parameter $\phi$ \ {(though they indeed depend on $M_i$, $M_f$)}, we shall now proceed to argue how the area of FZs (for the sector $n=0$) should look at these
critical rates  $\tau= \tau_c^1$ and $\tau_c^2$. Note that at $\tau=\tau_c^1$, $p_{\mathbf{k}}$  corresponding to
the mode $k_x=0,k_y=0$  becomes $1/2$ and it is {\it only} the FZ corresponding to   this particular point of the area which  touches the real time axis as shown in Fig.~\ref{tau1cfz}.  On the contrary, at $\tau= \tau_2^c$, the internal boundary of the area of FZs touches the real axis  for the mode $k_x= 2\pi/3, k_y=0$, in
addition to already existing  two  crossings due to the intermediate behaviour. In short, the area crosses the real time axis for two momenta and  additionally touches it  for the above TRIM point (Fig.~\ref{tauc2fz}).
When $\tau$ exceed $\tau_c^1$ ($\tau_c^2)$ even infinitesimally, the intermediate
(re-entrant) behaviour as shown in Figs.~\ref{twodptpk}-\ref{onedptfz} sets in. 
\begin{figure}[]
\centering
\subfigure[]{%
\includegraphics[width=.22\textwidth,height= 5cm]{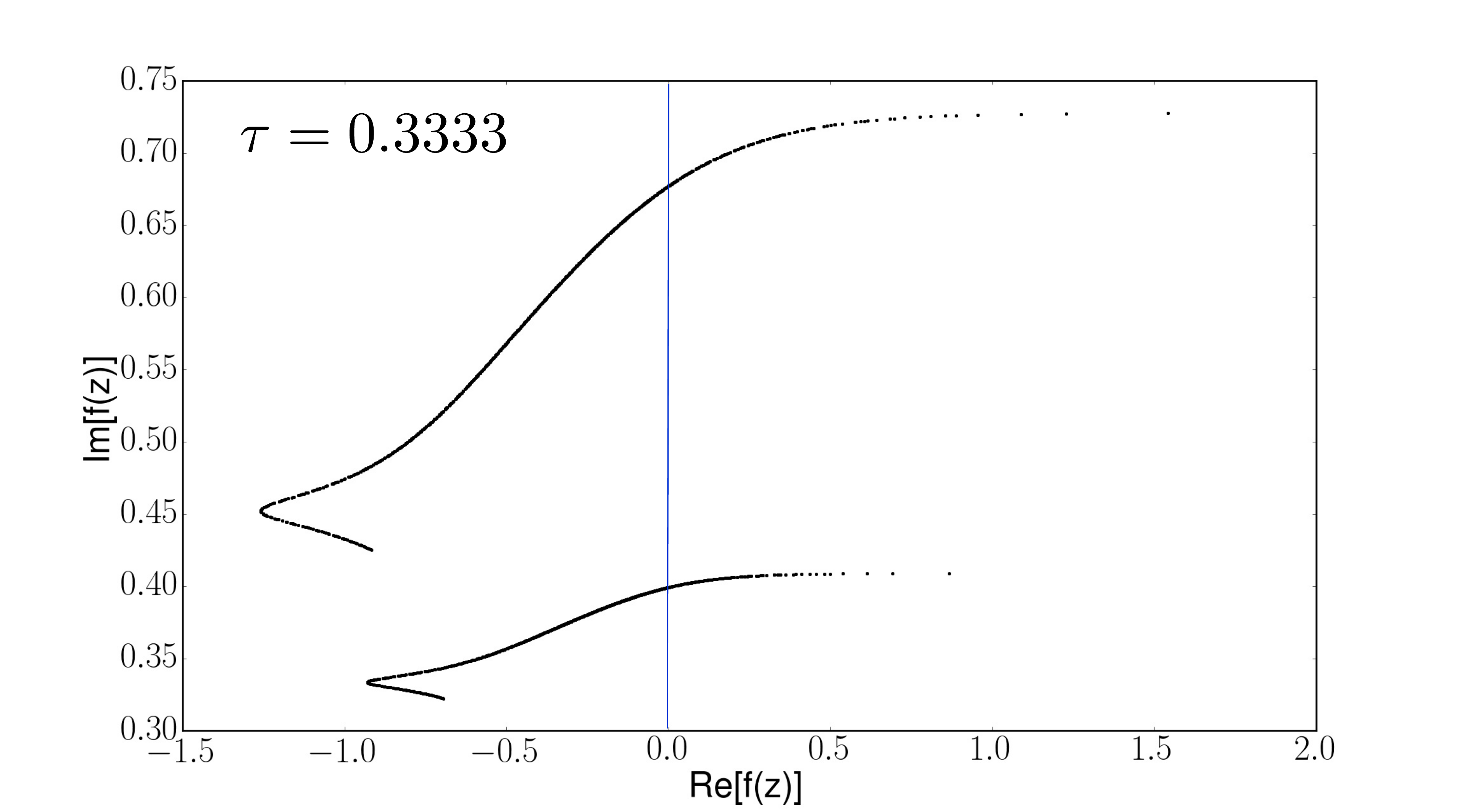}
\label{fig_2lines}}
\hfill
\quad
\subfigure[]{%
\includegraphics[width=.22\textwidth,height =5cm]{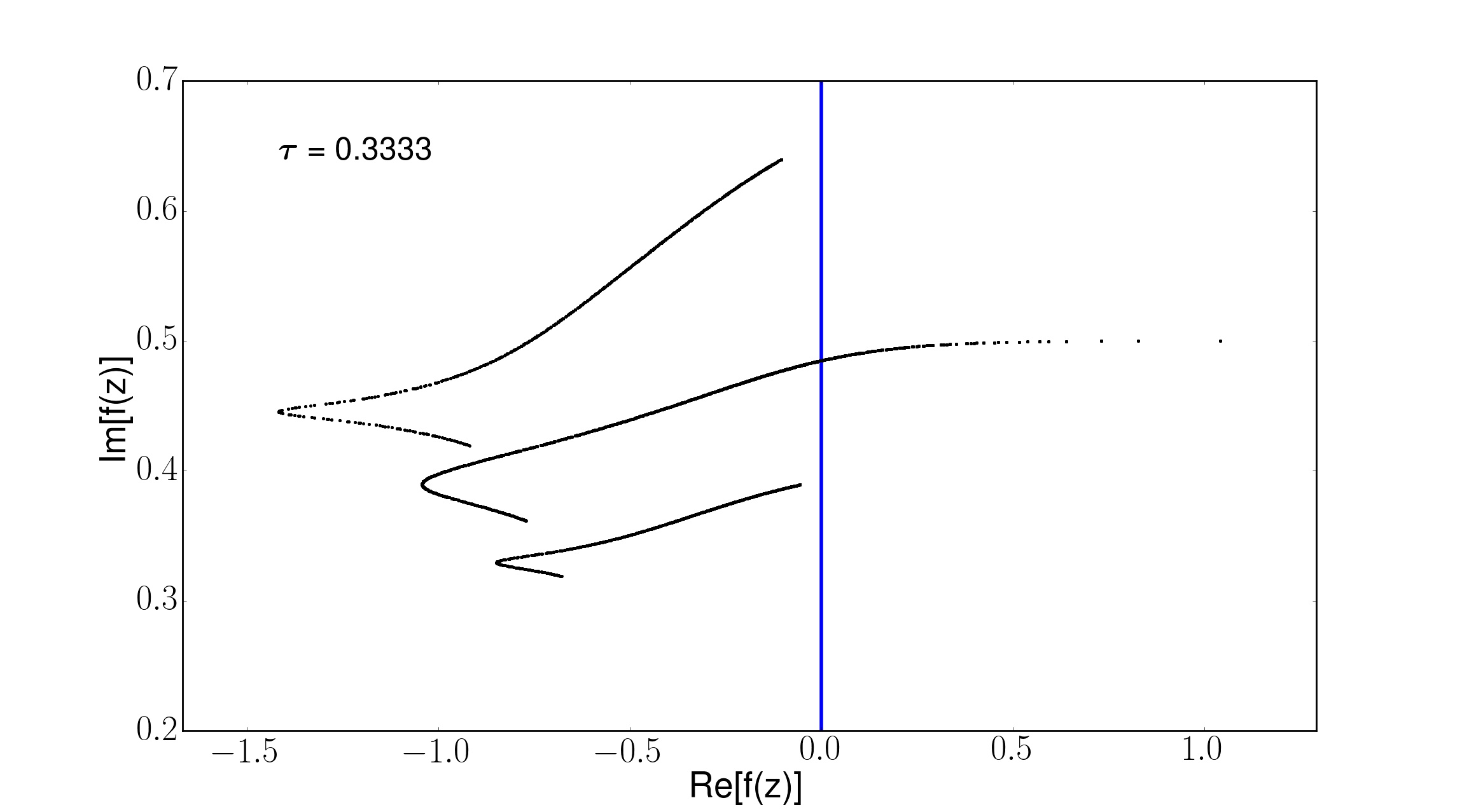}
\label{fig_3lines}}
\caption{(a) The FZs  when $M_H(\mathbf{k})$ assumes two constant values (with different signs) at two DPs; we obtain two disjoint lines
of FZs both crossing the real time axis at two different instants. (b) This refers to a hypothetical situation where the Haldane mass assumes three
constant values in three regions of the BZ as elaborated in the text and consequently one obtains three disjoint lines. }
\end{figure}

\section{Limiting situations: Formation of areas of FZs due to $M_H(\mathbf{k})$}

\label{sec_zero_MH}

We have already demonstrated that  the non-zero Haldane mass plays a
crucial role in generating areas of FZs in the complex $z$ plane. Furthermore, it vanishes for  the TRIM  points which determine the characteristic values of $\tau$, i.e., 
$\tau_1^c$ and $\tau_2^c$. In this section, we shall revisit the crucial role played by $M_H(\mathbf{k})$ by considering some limiting  situations. 
Let us first  analyze the situation when
the Haldane mass altogether vanishes, (i.e., the parameter $\phi=0$) so that the equilibrium model is simply a massive Dirac Hamiltonian
which does not possess any non-trivial topology {(see Fig.~\ref{fig_zero_phi}).}  \ {This model is well studied and known to mimic the graphene Hamiltonian on 
a Boron Nitride substrate.}  Although the results presented Fig.~\ref{fig_zero_phi} are obtained for the lattice model \ {i.e., Eq.~\eqref{eq_haldane_lattice} with $t_2=0$}, to get an intuitive understanding of the results,
we resort to  the  \ {low-energy} continuum limit: one can then recast the Hamiltonian} \eqref{eq_haldane_model} in the vicinity of DPs with $M_H(\mathbf{k})=0$, in the form:

\begin{equation}
\begin{aligned}
\mathcal{H} & = 
\begin{pmatrix}
M(t) &  k_x+i k_y \\
k_x-ik_y & - M(t)
\end{pmatrix},
\end{aligned}
\label{eq_haldane_model_continuum}
\end{equation}
where $k_x,k_y$ are measured from each DP.
The SM, $M(t)= -t/\tau$ is ramped from from $M_i=+3$  to $M_f=-3$ as shown in \ref{fig_haldane_phase} but with $\phi =0$; 
the system is thus quenched through two gapless DPs, which, in the present case, are time reversed partners of each other). Although  referring to Fig.~\ref{fig_phi01}, we find $p_{\mathbf{k}}=1$ at both DPs, FZs  (for the sector with $n=0$)   form  a  single  line in the $z$-plane
(Fig.~\ref{fig_phi02}) cutting the real time axis only once; the re-entrant behavior observed in Fig. \ref{twodptfz} also completely disappears { because the lines of FZs associated with the upper and lower
DPs fall on top of each other coalescing into the observed single line.}

To comprehend this
observation, let us note in the present case, one
can argue that the modes $\mathbf{k^*}$ (for which $p_{\mathbf{k^*}}=1/2$), satisfy the condition $|\mathbf{k^*}|^ 2= (k_x^*)^2 + (k_y^*)^2 ={\rm constant}$.
 More importantly, $e^2(\mathbf{k})=\sqrt{|\mathbf{k}|^ 2+M_f^2}$ is no-longer anisotropic, rather  depends only  on $|\mathbf{k}|$. Evidently, the  entire range of $\mathbf{k^*}$ leads to the same value of 
 $e^2(\mathbf{k^*})$.  Consequently, the system is effectively one-dimensional.
of real time  (for each $n$) at which Re[$f(t)$] itself shows a cusp singularity, as presented in Fig.~\ref{fig_phi03},
with a  discontinuous change in its first derivative.
Even for an infinitesimally small $\phi$, the areas of FZs reappear along with the re-entrant behavior (Fig.~\ref{fig_phi04}).

Let us now extend to the situation when the Haldane mass is $M_H^{\alpha}(\mathbf{k})$ in the continuum limit is positive at one DP and negative at the other.
Expanding around these DPs labelled by the index $\alpha=\pm 1$,  we  arrive at two Hamiltonians:
\begin{equation}
\begin{aligned}
\mathcal{H}^{\alpha} & = 
\begin{pmatrix}
M(t)+M_H^{\alpha} &  k_x+i k_y \\
k_x-ik_y & -(M(t)+M_H^{\alpha})
\end{pmatrix},
\end{aligned}
\label{eq_haldane_model_2line}
\end{equation}
where the Haldane mass  $M_H = -3 t_2 \alpha \sin \phi$. In this case, though $M_H$ is independent of the magnitude of the quasi-momentum, the two DPs sense different $M_H$. This is reflected in the behavior of FZs, which now form two disjoint
lines corresponding to the sector $n=0$ as shown in Fig. \ref{fig_2lines}, which cut the real axis at two different instants of time. 
We emphasise that an area of FZs does not appear in this case and \ {at the same time, there is a deviation from the one dimensional situation depicted in
Fig. \ref{fig_phi02}.}

The \ {continuum limit discussed} above can also be viewed as a special situation where $M_H$ is positive in the upper half of the BZ while negative in the lower half. This can be further extended
to an artificial situation where $M_H$ assumes three different values along three regions parallel to the $k_x$ -axis of the BZ where the central region extends from $k_y = \pi/\sqrt{3}$ to 
$k_y = - \pi/\sqrt{3}$ with a constant Haldane mass $M_H^{2}$, whereas the regions above and below it have constant masses  $M_H^1$ and $M_H^3$, respectively, with the condition $M_H^1>M_H^2>M_H^3$ \ {so that the topological structure presented in the Fig.~\ref{fig_haldane_phase} remain intact}. In this case, one finds three disjoint lines of FZs as shown in Fig.~\ref{fig_3lines}. When  generalized to  $n$ similar stripes with constant mass $M_H^{(n)}$ for the $n$-th stripe, {more and more} disjoint lines of FZs appear.  Equipped with this important  observation,  one can now view the continuous variation of the $M_H(\mathbf{k})$
across the BZ (Fig.~\ref{fig_haldane_mass}), as a limiting situation when area of  these regions becomes infinitesimal; in a such scenario these disjoint lines coalesce to form
a quasi-area which becomes a real dense area in the thermodynamic limit.  This provides an intuitive explanation of how the anisotropy in $M_H(\mathbf{k})$ generates areas of FZ
from otherwise disjoint lines. \ {In short, there is a deviation from the emergent one-dimensional behavior as soon as the topological mass term is incorporated in any form using  discrete variations over the BZ so that the topological structure remains invariant;  corresponding  disjoint lines of FZs (of the same sector) coalesce to form an area in the limit  when the Haldane mass is allowed to vary continuously over the BZ as happens in the lattice version of the model \eqref{eq_haldane_lattice}.}

\section{Experimental Connections}

\label{sec_expt}
Jotzu $et~al.$ \ct{jotzu14} demonstrated that the experimental realisation of the Haldane Hamiltonian is indeed possible with ultracold atoms in optical lattices periodically modulated in time. A rotating force, as proposed by Oka and Aoki \ct{oka09}, in a honeycomb lattice breaks TRS and leads to the necessary complex next-neighbor hopping giving rise to a Floquet Hamiltonian with Haldane-like mass term. In the above mentioned experiment \ct{jotzu14}, a honeycomb optical lattice, was created by several laser beams arranged in the $x-y$ plane, which generates the hopping terms $h_1$ and $h_2$ in Eq.~\eqref{eq_haldane_model}, and the atoms in A and B sublattices were separated via a large offset (or the  SM 
denoted by $M$  in $h_3$ of Eq.~\eqref{eq_haldane_model}), which can be tuned by changing the polarisation of the laser.  The lowest band was then filled with non-interacting, ultracold gas of fermionic $^{40}\rm{K}$ atoms before ramping up a sinusoidal modulation of the lattice position along the $ x$ and $y$ directions which resulted in TRS preserving linear ($\phi = 0,\pi,2\pi$), and TRS breaking circular ($\phi=\pi/2$) and elliptical ($\phi\neq 0,\pi/2,\pi,2\pi$) Haldane-like mass term $M_H$ given in the Haldane Hamiltonian. Later, Flaschner $et~al.$ \ct{flaschner16} using a similar setup studied the time evolution of fermionic quantum gases in such a hexagonal optical lattice after a rapid quench from a topologically trivial system into a Haldane-like system by quenching between a static and a Floquet Hamiltonian which generates the dynamics. The initial many-body ground state is a band insulating state in the lowest band of such a hexagonal lattice with a large offset between the A and B sites. A quench into the final Floquet Hamiltonian (manifested through resonant circular lattice shaking) was then performed. Finally, a momentum and time-resolved state tomography measured this out of equilibrium dynamics after the stroboscopic evolution times of the Floquet Hamiltonian, yielding the evolution of the many-body state on the Bloch sphere, thereby, ascertaining the occurrence of DQPTs. Similarly, a slow ramping instead of a rapid quench into the final Floquet Hamiltonian seems plausible and may indeed corroborate our results. It should also be noted that the finite time LZ limit, on which we focus on, is more feasible to experimental verification than an infinite LZ limit. Furthermore, there has also been a recent work in which Yang-Lee zeros have been observed experimentally \ct{peng16}.

\section{Conclusion}

\label{sec_conclusion}

Exploiting the two band nature of the topological Haldane model which enables us to employ extensive investigations of  finite time LZ problems, we establish
a deep connection between the equilibrium  topology of the Haldane model and subsequent DQPTs following a linear ramping of the SM from one non-topological phase to the other. Other than the re-entrance of areas of FZs in the extreme slow limit which is an artefact of quenching across two DPs,
we establish the existence of an intermediate range of the quenching rate for which areas of FZs  cross the real axis  only once leading to  two (instead of four as in the re-entrant case)  instants of non-analyticities 
in Re$[f(t')]$ for a fixed value of $n$. This intermediate region, which do not show up in the case of the transverse Ising chain \ct{sharma16}, exists only because of the position of the DPs inside the BZ;  this   allows $p_{\mathbf{k}}$ to vanish above (below) the upper (lower) DP along the line $k_y=2\pi/3$ while keeping it always  greater than $1/2$ between them for appropriate values of $\tau$.

The quasi-momentum dependent topological Haldane mass, 
plays the most crucial role:
 it is the presence of a non-zero $M_H(\mathbf{k})$ that results in the appearance of areas of FZs in the complex $z$ plane as well as the re-entrant behavior in the extreme
 slow limit. When $M_H(\mathbf{k})$ is switched off, the areas shrink to lines of FZs giving
 rise to an emergent one dimensional behavior with non-analyticities in Re$[f(t')]$ itself (and discontinuities in Re$[f'(t')]$) at those instants  when these lines cut the real axis; furthermore, the re-entrant behavior completely disappears even in the extreme slow  limit. 
 {Focussing
 on the re-entrant situation, if the
 value of $M_H(\mathbf{k})$ 
 is slowly reduced, the areas corresponding to two branches of a lobe (for a given $n$) become thinner and thinner approaching two lines which eventually
 fall on top of each other when $M_H(\mathbf{k})=0$. 
  We have also illustrated by considering hypothetical situations when the BZ is divided into regions with different (but constant within a region) $M_H$, how otherwise disjoint lines of Fisher zeros coalesce into dense areas in the real model 
with continuously varying $M_H$   in the thermodynamic limit. Crucially, the critical rates  $\tau_1^c$
 and $\tau_2^c$, marking the  crossover between the no-DQPT, intermediate and re-entrant behavior of FZs are solely determined by the TRIM point
 included in the BZ for which $M_H =0$, thereby, further emphasising the immense importance of the Haldane mass in dictating the nature of DQPTs.

  
 
  {A pertinent question that may arise at this point that how general the conclusion drawn based on the symmetry and topology will be, more because of the fact that numerical results
 are presented for a particular quenching protocol.  Out of the three possible behaviour of DQPTs,
 what situation will be encountered depends upon the quenching amplitude for sudden quenching and the rate of quenching for the slow quenching case (between finite values of $M_i$ and $M_f$). 
 But what is robust is that   there will be areas of FZs only when  the Haldane mass  has a continuous variation across the Brillouin zone for the present model. If it is altogether absent
 there will be lines of FZs.  On the contrary, when it has a discrete variation over the BZ  keeping the topological structure intact, there will be a deviation from the one-dimensional behaviour. These conclusions will
 hold true irrespective of the fact whether the quenching is sudden or slow or whether the DQPTs exist or not in the context of the model considered here. }
 
   {Secondly, in the slow ramping case of the Haldane model, the crossover rates $\tau_c^1$
 and $\tau_2^c$ will always be determined by the excitation probabilities at the corner TRIM point ($k_x=k_y=0$) and the central TRIM point of the BZ; the Haldane mass being zero
 at these points, the value of $\tau_c^1$ and $\tau_c^2$ are independent of the parameter $\phi$ for any-nonzero $\phi$: to be precise, when $p_{\mathbf k}$ both at the corner TRIM point ($k_x=k_y=0$) as well as at the central TRIM point exceed $0.5$, there will be a no-DQPT behaviour while both these fall below $0.5$, there will be an emergent (two-DQPTs)
 behaviour. Finally when $p_{\mathbf k} <0.5$  at the corner point and exceeds 0.5 at the central point, there will be intermediate (one-DQPT) behaviour. In fact, the above argument
 would hold true for sudden quenches ($\tau =0$) also; for example, one can conceive a sudden quenching scheme of the SM with $M_i=2.6$ to $M_f=-2.6$; there will be an intermediate behaviour of FZs and hence
one-DQPT.   Analysing the corresponding analogue of Fig. (4c), one would find that the above arguments  would explain the intermediate behaviour observed in 
this case. In summary,  the observations concerning the importance of the Haldane mass and the importance of TRIM points in determining the occurrence
of three different regions   are robust and
 will be able to explain all the numerical findings in the context of the Haldane model. }
}
 
 {Let us conclude with a few clarifying comments: (i) The graphene Hamiltonian with the SM (which has similarly been quenched) although anisotropic in $k_x$, $k_y$ away from the DPs is still topologically trivial and does not lead
 to the generation of areas of FZs, essentially requiring the topological Haldane mass term to generate the areas. 
  (ii) Furthermore, the addition of a hypothetical isotropic in $k_x$ and $k_y$ mass term such as $\cos (k_x^2 +k_y^2)$ alongwith the SM to the graphene Hamiltonian may generate areas of FZs in the complex $z$-plane. But such an isotropic mass term is non-topological. Since,  our focus in this paper is strictly limited to the study of  the topological Haldane model, where the quasi-momentum dependent Haldane mass term plays crucial role both in equilibrium
 topology and also in DQPTs, we do not consider
 such possibilities in this work.
 (iii) For slow quenching across a single critical point, when the $M_H$ is present, the areas of FZs obviously show either no-DQPT or one-DQPT behaviour as has been shown in the sudden quenches case in \ct{vajna15} whereas when $M_H$ is zero, here too FZs form lines.
  (iv)  Finally, we would like to emphasize that in the infinite LZ limit ($M_i =-M_f \to \infty$), it can be argued that the effect of topology gets effectively wiped out.}

{Given the recent experimental observation of DQPTs as discussed in Sec. \ref{sec_expt},    our work, naturally, opens up the possibility of further research in several new directions. One such example would be to explore the role of equilibrium topology in determining DQPTs in higher dimensional quenched topological models. Another intriguing question that inevitably needs to be addressed would be the effect of the edge states appearing in equilibrium topological models with open boundary condition, on the DQPTs described here.}

 \section*{Acknowledgements}
We acknowledge interesting discussions with Shraddha Sharma. We also thankfully acknowledge G. Baskaran and A. Polkovnikov for critical  comments. 
 AD acknowledges financial support from SERB,
DST, India. 


\appendix
 \section{A brief note on the Haldane model}
 
\label{app_model}

\begin{figure}
\begin{center}
\includegraphics[width=0.4\textwidth,height=0.3\textheight]{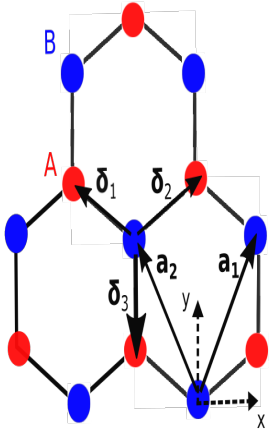}
\end{center}
\caption{A hexagonal lattice on which the topological Haldane model resides with $\mathbf{a_1}$ and $\mathbf{a_2}$ as the lattice vectors.}
\label{latt}
\end{figure}

We consider a 2D model on a hexagonal lattice comprised of two triangular sublattices $A$ and $B$ as shown in Figure \ref{latt}. It can be understood as the composition of a nearest neighbour tight binding graphene-like Hamiltonian along with a Haldane mass ($M_H$) and sublattice symmetry (SLS) breaking Semenoff  mass ($M$) term, with the Hamiltonian 
\begin{equation}
\begin{aligned}
&\mathcal{H} =  t_1 \sum_{i,j=N1}  \left(c^\dagger_{iA} c_{jB} + h.c. \right) \\
&+  t_2 \sum_{i,j=N2}  \left( e^{i\phi_{ij}} \left(c^\dagger_{iA} c_{jA} + c^\dagger_{iB} c_{jB} \right) + h.c. \right) \\
&+ M \sum_{i\in A} \hat{n}_i - M \sum_{i\in B} \hat{n}_i.
\end{aligned}
\label{eq_haldane_lattice}
\end{equation}
The $c_{iA(B)}$s are spinless fermionic operators on sublattice $A$ ($B$), and the $t_2$s are the next  nearest neighbour hopping interaction strengths. The time reversal symmetry  of this model is broken by the phase factor $\phi_{ij}=\pm \phi$, originating from the staggered magnetic field and is positive for anticlockwise next nearest neighbour hopping.

Fourier transforming into k-space the Hamiltonian becomes
\begin{equation}
\begin{aligned}
\mathcal{H} & = \begin{pmatrix}
c^{\dagger}_A(\mathbf{k}) & c^{\dagger}_B(\mathbf{k}) 
\end{pmatrix}
\mathbf{h} (\mathbf{k})
\begin{pmatrix}
c_A(\mathbf{k}) \\
c_B(\mathbf{k}) 
\end{pmatrix},
\end{aligned}
\end{equation}
where 
\begin{equation}
\mathbf{h}(\mathbf{k}) = \sum^3_{i=0} h_i (\mathbf{k}) \sigma_i.
\label{eq_ham}
\end{equation}
The $\sigma_{i}$, $i \in \left\{ 1,2,3\right\}$ are the Pauli matrices, $\sigma_0$ is the identity matrix and $a$ is the lattice constant. 
We have 
\begin{align}
h_0 &= 2 t_2 \cos (\phi) \biggl[\cos \left( \mathbf{k} \cdot \mathbf{a}_1 \right) + \cos \left( \mathbf{k} \cdot \mathbf{a}_2 \right) \nonumber \\
&\hspace{4mm} + \cos \left( \mathbf{k} \cdot \left(\mathbf{a}_1 - \mathbf{a}_2 \right)\right) \biggr], \\
h_1 &= t_1 \biggl[ 1 + \cos \left( \mathbf{k} \cdot \mathbf{a}_1\right) + \cos \left(\mathbf{k} \cdot  \mathbf{a}_2 \right) \biggr]\nonumber\\
h_2  &= t_1   \biggl[\sin  \left( \mathbf{k} \cdot \mathbf{a}_1 \right) +\sin \left( \mathbf{k} \cdot\mathbf{a}_2 \right) \biggr],\\
h_3 & = M + M_H, \\
M_H &= 2 t_2 \sin (\phi) \biggl[\sin  \left(\mathbf{k} \cdot \mathbf{a}_2 \right)  - \sin \left( \mathbf{k} \cdot \mathbf{a}_1 \right) \nonumber\\
&\hspace{4mm}+ \sin \left( \mathbf{k} \cdot \left(\mathbf{a}_1 - \mathbf{a}_2 \right) \right)\biggr],
\end{align}
where $\mathbf{a}_1 =\frac{a}{2} \left(3, \sqrt{3}, \right)$ and $\mathbf{a}_2 =\frac{a}{2} \left(-3, \sqrt{3} \right ) $ as shown in Figure \ref{latt}.

%
%
%
%

\end{document}